\documentclass[aps,prl,twocolumn,floatfix,superscriptaddress,showpacs,longbibliography]{revtex4-1}
\usepackage{xfrac}
\usepackage{amsfonts}
\usepackage{dsfont}
\usepackage{amsmath}
\usepackage{amssymb}
\usepackage{mathtools}
\usepackage{graphicx}
\usepackage{color}
\usepackage{bbm}
\usepackage[normalem]{ulem}
\usepackage[hidelinks]{hyperref}
\hypersetup{
     colorlinks   = true,
     citecolor    = blue,
     linkcolor    = blue
}
\usepackage{subfigure}
\usepackage{mathptmx}
\usepackage{times}

\newcommand{\tr}{{\rm tr}}

\newcommand{\ignore}[1]{}
\newcommand{\nobibentry}[1]{{\let\nocite\ignore\bibentry{#1}}}

\newcommand{\bibfnamefont}[1]{#1}
\newcommand{\bibnamefont}[1]{#1}
\newcommand{\qav}[1]{\left<#1\right>}

\DeclareMathAlphabet{\mathcal}{OMS}{cmsy}{m}{n}

\begin{document}

\title{Dynamically Induced Heat Rectification in Quantum Systems}

\author{Andreu Riera-Campeny}\affiliation{F\'{\i}sica Te\`{o}rica: Informaci\'{o} i Fen\`{o}mens Quantics. Departament de F\'{\i}sica, Universitat Aut\`{o}noma de Barcelona, 08193 Bellaterra, Spain}
\author{Mohammad Mehboudi}
\affiliation{ICFO-Institut de Ci\`{e}ncies Fot\`{o}niques, The Barcelona Institute of
	Science and Technology, 08860 Castelldefels, Spain.}
\author{Marisa Pons}
\affiliation{Departmento de F\'isica Aplicada I, Universidad del Pa\'is Vasco, UPV-EHU, Bilbao, Spain.}
\author{Anna Sanpera}
\affiliation{F\'{\i}sica Te\`{o}rica: Informaci\'{o} i Fen\`{o}mens Quantics. Departament de F\'{\i}sica, Universitat Aut\`{o}noma de Barcelona, 08193 Bellaterra, Spain}

\affiliation{ICREA, Psg. Llu\' is Companys 23, 08001 Barcelona, Spain.}

\begin{abstract}
Heat rectifiers are systems that conduct heat asymmetrically for forward and reversed temperature gradients. We present an analytical study of heat rectification in linear quantum systems. We demonstrate that asymmetric heat currents can be induced in a linear system only if it is dynamically driven. The rectification can be further enhanced, even achieving maximal performance, by detuning the oscillators of the driven network. Finally, we demonstrate the feasibility of such driven harmonic network to work as a thermal transistor, quantifying its efficiency through the dynamical amplification factor.
\end{abstract}		

\date{\today}
\maketitle


Rectifiers are physical systems capable of conducting energy asymmetrically ---whether electric, magnetic, thermal...--- and are an essential building block in many technological applications. Although thermal rectifiers are crucial components to manipulate heat currents and construct phononic devices, so far no efficient and feasible thermal diodes have been found. Such device, when connected to two thermal baths at different temperatures, conducts heat asymmetrically if the temperatures of the baths are interchanged. This effect allows for an effective heat dissipation with a suppressed backflow reaction. 

To date, most theoretical proposals on classical heat rectifiers (see \cite{RevModPhys.84.1045} and references therein) have been based either on the use of inohomogenous materials \cite{Chang1121,Benenti2016,PhysRevLett.88.094302,PhysRevLett.108.234301,refId0,liu2014important} exploiting nonlinear interactions, or doping the systems with impurities while remaining in the linear regime~\cite{refId009}. Also, the feasibility of microscopic systems acting as thermal devices has been recently addressed in, for instance, phononic refrigerators in the classical \cite{Arrachea2012} and quantum \cite{Freitas2017} regimes, or heat rectifiers in diferent platforms: quantum dots \cite{1367-2630-10-8-083016}, nonlinear solid-state quantum circuits~\cite{PhysRevB.79.144306}, few-level systems~\cite{Joulain2016, 2018arXiv180604794W}, or hybrid configurations~\cite{PhysRevE.80.041103}.

Here, we address analytically and in full generality heat rectification in quantum systems under generic linear interactions. To this aim, we assume a network of harmonic oscillators coupled to two thermal reservoirs and investigate how asymmetric heat fluxes can be induced in such setup. First, we revisit the static scenario showing that  linearity forbids heat rectification, regardless of any asymmetry in the harmonic network or in its coupling with the baths. Second, we demonstrate that heat rectification in a linear quantum system is possible if the system is periodically driven. This is our main result. Such feature is a consequence of two facts: (i) injecting/extracting work into/from a system by an external agent is a useful resource to redistribute energy and (ii) by periodically driven a system, new asymmetric heat transport processes --that have no analog in static scenarios-- are induced. By using the Floquet formalism we identify precisely the quantum processes leading to heat rectification. Finally, we also demonstrate the suitability of driven harmonic networks as heat transistors.

\begin{figure}[h]
	\includegraphics[width=1\linewidth]{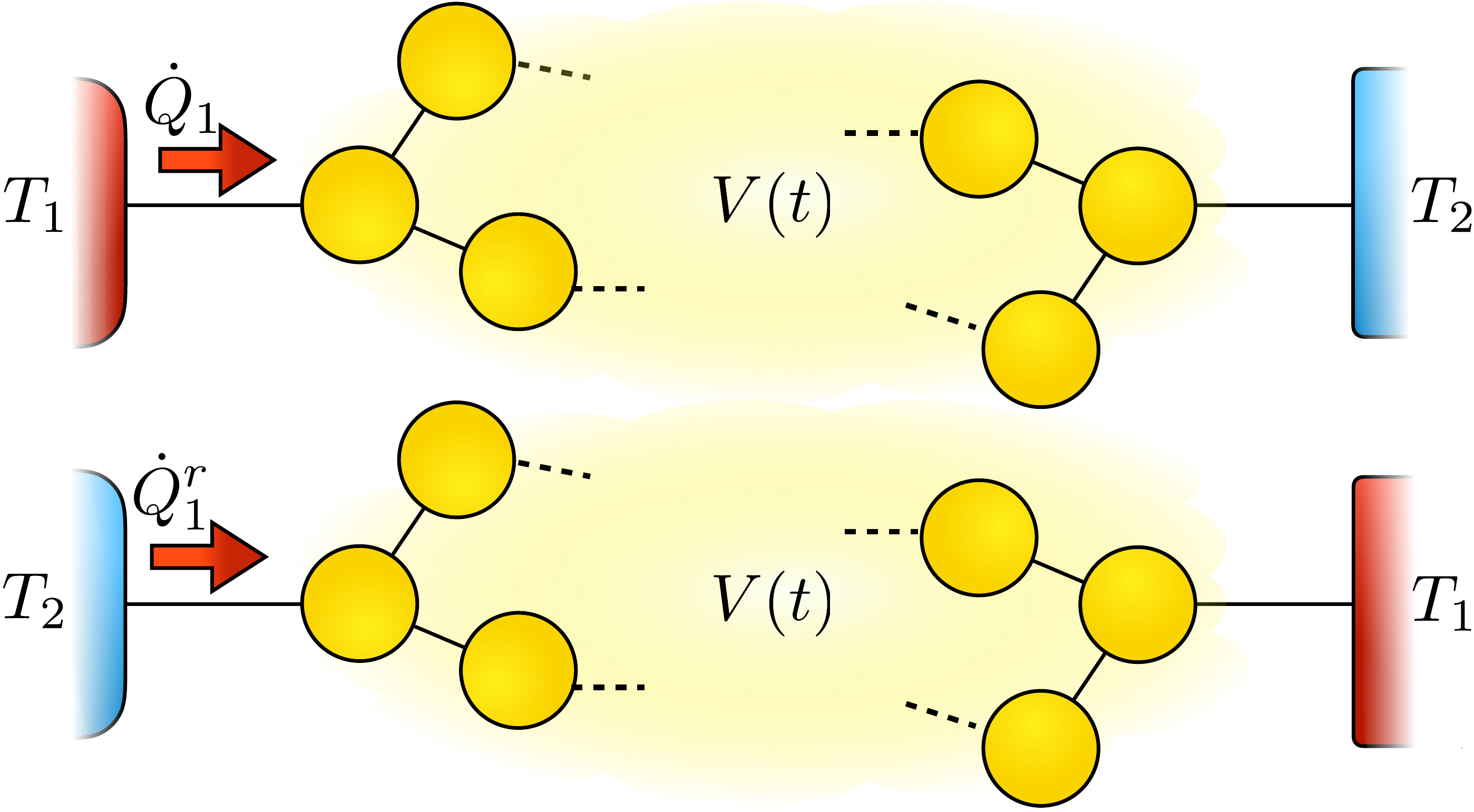}
	\caption{Sketch of a heat rectification setup where a system $S$ with linear interactions $V(t)$ is connected to two reservoirs at fixed temperatures $T_1$ and $T_2$. (Top) forward configuration, (bottom) reversed configuration. }
	\label{fig:setup}
\end{figure}

Consider a system $S$ coupled to two reservoirs at fixed temperatures $T_1$ and $T_2$. In any out-of-equilibrium scenario, the energy of the reservoirs is redistributed through the system in the form of heat currents. The system $S$ works as a \textit{heat rectifier} if the magnitude of the heat current depends on the sign of the temperature gradient~\cite{Roberts2011}. Such asymmetric heat current is conventionally quantified by the {\it rectification coefficient} \cite{PhysRevB.79.144306,Joulain2016}:
\begin{align}
R(\dot{Q}_1,\dot{Q}_1^r) \coloneqq \frac{|\dot{Q}_1+\dot{Q}_1^r|}{\text{max}(|\dot{Q}_1|,|\dot{Q}_1^r|)},\label{eq:rectification_coef}
\end{align}
where $\dot{Q}_1$ ($\dot{Q}_1^r$) is the heat flowing into the system from the first reservoir in the forward (reversed) configuration. For $T_1 \geq T_2$, forward and reversed configurations refer to, respectively, negative and positive temperature gradients (see Fig.~\ref{fig:setup}). Notice that $0 \leq R(\dot{Q}_1,\dot{Q}_1^r) \leq 2$. The lower bound is achieved for a system that conducts symmetrically: $\dot{Q}_1^r=-\dot{Q}_1$, while the upper bound is saturated for heat fluxes that are independent of the sign of the temperature gradient:  $\dot{Q}_1 = \dot{Q}_1^r$. 
A system that blocks the heat flux in either configuration fulfills $R(\dot{Q}_1,0)=R(0,\dot{Q}_1^r)=1$. 


To provide exact closed expressions for the rectification coefficient we consider the system S to be a network of $N$ linearly-coupled harmonic oscillators with time-dependent Hamiltonian: 
\begin{align}
H_S(t) =  \frac{P^TM^{-1}P}{2}+\frac{X^T V(t) X}{2}. 
\end{align}
Here, $X = (x_1, \cdots, x_N)^T$ and $P=(p_1, \cdots, p_N)^T$ are the vectors of position and momentum operators of the network---fulfilling the standard bosonic algebra $[x_i,p_j]=\text{i}\hbar\delta_{ij}$. The matrix $M = \text{diag}(m_1, \cdots, m_N)$ contains the masses of the oscillators of the network. The diagonal part of $V(t)$ captures the oscillators self-energy, whereas its off-diagonal elements encode the coupling between the positions of different oscillators. We assume that the system S is coupled to a set of reservoirs, being $H_{R_\alpha}$ the Hamiltonian of the reservoir $\alpha$. Each reservoir is described by a collection of non-interacting bosonic modes $\mu$, with Hamiltonian
\begin{align}
H_{R_\alpha} =\frac{P_{\alpha}^{T} M_{\alpha}^{-1} P_{\alpha}}{2} + \frac{X_{\alpha}^{T}M_{\alpha}\Omega_{\alpha}^{2}X_{\alpha}}{2}, \label{eq:Hbath}
\end{align}
where $X_\alpha = (\{x_{\alpha,\mu}\})^T$ and $P_\alpha =(\{p_{\alpha,\mu}\})^T$ are the reservoir position and momentum vector operators respectively. The masses and frequencies of the modes of the reservoir $\alpha$ are given by $M_\alpha = \text{diag}(\{m_{\alpha,\mu}\})$ and $\Omega_{\alpha} =\text{diag}(\{\omega_{\alpha,\mu}\})$ respectively. The interaction Hamiltonian between the system S and the reservoir $\alpha$ is 
\begin{align}
&H_{SR_\alpha} = - X^T C_{\alpha} X_{\alpha}, \label{eq:Hsysbath}
\end{align}
where the rectangular matrix $C_\alpha$ encodes the coupling between the system and the reservoir $\alpha$.  Couplings of the form $H'_{SR_\alpha} =-X^T \tilde{C}_\alpha P_\alpha$, that appear frequently in quantum optics, can be treated with the same techniques we display here. The total Hamiltonian is given by $H= H_S(t) + \sum_\alpha H_{SR_\alpha} + \sum_\alpha H_{R_\alpha}$.

Under the natural assumption that the system is initially uncorrelated from the reservoirs  i.e., $\rho(t_0) = \rho_S(t_0) \bigotimes_\alpha \rho_{R_\alpha}(t_0)$, the equations of motion for $\{X_\alpha\}$ and $\{P_\alpha\}$  can be solved leading to the quantum Langeving equation that describes the steady state of system \cite{Grabert1984,Breuer2007}:
\begin{align}
M\ddot{X} + V(t)X - \chi(t)\star X = B(t), \label{eq:qle}
\end{align}
where $\star$ denotes the convolution. More details of the derivation are given in the Supplemental Material. 
The {\it susceptibility matrix} $\chi(t)$ acts as a damping source, whereas the {\it noise vector} $B(t)$ as an external force. Their explicit expressions are
\begin{align}
\chi(t) &= \theta(t) \sum_\alpha C_\alpha (M_\alpha \Omega_\alpha)^{-1} \sin(\Omega_\alpha t) C^T_\alpha,\label{eq:susceptibility}\\
B(t) &= \lim_{t_0 \to -\infty} \sum_\alpha C_\alpha \left[\sin(\Omega_\alpha(t-t_0))X_\alpha(t_0)+\right.\nonumber\\
&\hspace{0.25 in} +\left.(M_\alpha\Omega_\alpha)^{-1}\cos(\Omega_\alpha(t-t_0))P_\alpha(t_0)\right],\label{eq:noise_vector}
\end{align}
with $\theta(t)$ being the Heaviside step function. The dissipation kernel $\eta(t)$, widely used in the literature, is related to the susceptibility: $\chi(t) = \theta(t) \eta(t)$. We also introduce the spectral density $J(\omega) =\sum_\alpha J_\alpha(\omega) = 1/2\sum_\alpha C_\alpha (M_\alpha \Omega_\alpha)^{-1} \Delta_\alpha(\omega) C_\alpha^T$, where $\Delta_{\alpha,\mu\nu}(\omega) = \delta_{\mu\nu}\delta(\omega-\omega_{\alpha,\mu})$. We further assume that all oscillators of the network are at most coupled to one reservoir. Therefore, the support of each $J_\alpha(\omega)$ is in an orthogonal subspace. One can think of $J_\alpha(\omega)$ as a measure of the number of modes in reservoir $\alpha$ whose frequencies lie between $\omega$ and $\omega+\text{d}\omega$. With the help of the spectral density it is easy to see $\chi(t) = \theta(t) 2\int_0^\infty\text{d}\omega J(\omega) \sin(\omega t)$. 
By extending $J(\omega)$ to negative frequencies, as $J(-\omega) \coloneqq -J(\omega)$, it follows that $\text{Im}\chi(\omega) = \pi J(\omega)$, which is a compact form of the {\it fluctuation-dissipation theorem} \cite{Kubo1966}. In what follows, we further assume that $\rho_{R_\alpha}(t_0) = \exp(-H_{R_\alpha}/k_B T_\alpha)/\text{tr}[\exp(-H_{R_\alpha}/k_B T_\alpha)]$ is a thermal state at temperature $T_\alpha$.

Let us now review the thermodynamical quantitites of interest. For a general open quantum system, a consistent definition for heat an work was presented in \cite{Esposito2010}. The work rate is given by $\dot{W}= \qav{\partial_t H}$ where $\qav{\cdots}$ denotes the average over the total density matrix $\rho$ . If we assume that only $H_S(t)$ varies with time, $\dot{W} =  \qav{\partial_t H_S(t)}$. The heat current from the reservoir $\alpha$ to the system S is given by $\dot{Q}'_\alpha = -\text{d} \qav{H_{R_\alpha}}/\text{d}t $. Notice that these definitions of heat currents and work rate are completely general and not limited to any specific model. Alternatively, one can define the heat rate as $\dot{Q}_\alpha= \text{i}/\hbar \langle [H_{SR_\alpha}, H]\rangle\coloneqq\text{i}/\hbar \langle [H_{SR_\alpha}, H_S] \rangle$. Although in general both definitions do not coincide, for time-averaged linear systems at the steady-state it is easy to show that $\bar{\dot{Q}}_\alpha = \bar{\dot{Q}}'_\alpha$ (see \cite{Freitas2017}). The energy of the system S changes at rate: 
\begin{align}
 \frac{\text{d}}{\text{d}t}\left\langle H_S \right\rangle = \sum_\alpha\frac{\text{i}}{\hbar}\left\langle [H_{SR_\alpha}, H_S] \right\rangle + \left\langle \frac{\partial}{\partial t} H_S\right\rangle = \sum_\alpha \dot{Q}_\alpha + \dot{W}, \label{eq:change_energy}
\end{align} 
which has the form of the first law of thermodynamics.

We first revisit the static network, where $V(t)=V_0$ encodes a completely general time-independent linear interaction between the nodes of the network and, therefore, $\dot{W} = 0$. Since in steady state $\text{d} \langle H_S \rangle/\text{dt} =0$, the first law reduces to $\sum_\alpha \dot{Q}_\alpha = 0$. The steady-state solution for $X$ can be found using the Green's function technique \cite{Dhar2006}, leading to the following expression for the heat currents
\begin{align}
\dot{Q}_\alpha &= \sum_{\beta \neq \alpha} \int_\mathbb{R} \text{d}\omega \mathcal{T}^{0}_{\alpha\beta}(\omega)\left(n_\alpha(\omega) -n_\beta(\omega)\right),\label{eq:static_current_final}
\end{align}
where $n_\alpha(\omega) = (\exp(\hbar\omega/k_B T_\alpha) - 1)^{-1}$ is the thermal occupation number, $\mathcal{T}^{0}_{\alpha\beta}(\omega) = \hbar \omega \pi\text{tr}\left[J_\alpha(\omega)G_0(\omega)J_\beta(\omega)G_0^\dagger(\omega)\right]$ is the static heat transfer matrix, and $G_0(\omega)= \left(-\omega^2 M + V_0 + \chi(\omega)\right)^{-1}$ is the static Green's function in spectral domain.  Notice that, for the static case, $ \mathcal{T}^{0}_{\alpha\beta}(\omega) = \mathcal{T}^{0}_{\beta\alpha}(\omega)$ is a temperature-independent symmetric matrix. Particularly, in a scenario with two reservoirs, the forward and reversed currents necessarily fulfill $\dot{Q}_1 = -\dot{Q}_1^r$, hence $R(\dot{Q}_1,\dot{Q}_1^r)=0$. This is true for any interaction $V_0$ or any specific form of the spectral densities $J_{\alpha}(\omega)$. This seemingly surprising result can be explained by focusing on the normal-mode picture: A network of $N$-interacting oscillators can be mapped onto a system of $N$-independent oscillators---that is, the normal-mode representation. Each independent oscillator transfers heat symmetrically due to its invariance under reflection symmetry. 

Is it possible to attain asymmetric heat fluxes within harmonic systems? The answer is positive, but it requires exploring time-dependent systems. To this aim, we consider a periodic driving potential;  $V(t+\tau) = V(t)$ with $\tau$ being the period. Then, for the periodically driven system, all the terms in Eq.~\eqref{eq:change_energy} are time-dependent. However, in the steady state the averages over a period must remain constant and consequently the first law reads $0 = \sum_\alpha \bar{\dot{Q}}_\alpha + \bar{\dot{W}}$, where for any $O(t)$, its time average over a period is $\bar{O} \coloneqq \tau^{-1}\int_{t}^{t+\tau}\text{d}t' O(t')$. The existence of the steady state is not guaranteed in time-driven systems, for certain range of parameters, instabilities appear even in the case of a single oscillator. Assuming it exists and following \cite{PhysRevA.73.052311,Freitas2017,Freitas2018}, we use the Floquet theory of periodic differential equations to solve the dynamics of the system. 

We denote by $G(t,t')$ the Green's function associated to Eq.~\eqref{eq:qle}. Even though $G(t,t')$ is not necessarily periodic, one can define $P(t,\omega) = \int_\mathbb{R} \text{d}t' G(t,t') e^{\text{i} \omega (t-t')}$ which obeys $P(t+\tau,\omega) = P(t,\omega)$. Thus, one can Fourier expand 
\begin{align}
V(t) &= \sum_k V_k e^{\text{i} k \omega_d t} ,\label{eq:pot_expansion}\\
P(t,\omega) &= \sum_k A_k(\omega) e^{\text{i} k \omega_d t},\label{eq:p_expansion}
\end{align}
being $\omega_d=2\pi/\tau$. By substituting into Eq.~\eqref{eq:qle}, it follows that
\begin{align}
G_0^{-1}(\omega-k\omega_d)A_k(\omega)+\sum_{j\neq 0} V_j A_{k-j}(\omega) = \delta_{k,0}\mathds{1}.\label{eq:eom_amplitudesk}
\end{align}
With this at hand, we proceed to find the steady state averaged work rate and heat currents. They read as follows 
\begin{align}
&\bar{\dot{W}} = \sum_\alpha \int_\mathbb{R}\text{d}\omega \tilde{\mathcal{T}}_\alpha (\omega)\left(n_\alpha(\omega)+\frac{1}{2}\right),\label{eq:averaged_work}\\
&\bar{\dot{Q}}_\alpha = -\int_\mathbb{R}\text{d}\omega \tilde{\mathcal{T}}_\alpha(\omega) \left(n_\alpha(\omega)+\frac{1}{2}\right) +  \label{eq:averaged_current}\\
& \hspace{-5 pt}\sum_{\beta \neq \alpha} \int_\mathbb{R} \text{d}\omega\left[ \mathcal{T}_{\beta\alpha}(\omega)\left(n_\alpha(\omega)+\frac{1}{2}\right) - \mathcal{T}_{\alpha\beta}(\omega)\left(n_\beta(\omega)+\frac{1}{2}\right)\right],\nonumber
\end{align}
where, $\mathcal{T}_{\alpha\beta}(\omega)$ is now the dynamical heat-transfer matrix, whose off-diagonal elements are given by 
\begin{align}
\mathcal{T}_{\alpha\beta}(\omega) &= \sum_k \hbar (\omega-k\omega_d) r_{\alpha\beta}^k(\omega),\label{eq:q_alphabeta}
\end{align}
being $r_{\alpha\beta}^k(\omega) = \pi \text{tr}[J_\alpha(\omega-k\omega_d)A_k(\omega)J_\beta(\omega)A_k^\dagger(\omega)]$ the process rate, while $\tilde{\mathcal{T}}_\alpha(\omega) = \sum_\beta \mathcal{T}_{\beta\alpha}(\omega)$ is
\begin{align}
\tilde{\mathcal{T}}_\alpha(\omega) &= -\sum_k \hbar k \omega_d \pi\text{tr}\left[J(\omega-k\omega_d)A_k(\omega)J_\alpha(\omega)A_k^\dagger(\omega)\right].\label{eq:q_tilde}
\end{align}

\begin{figure}[t]
	\includegraphics[width=1\linewidth]{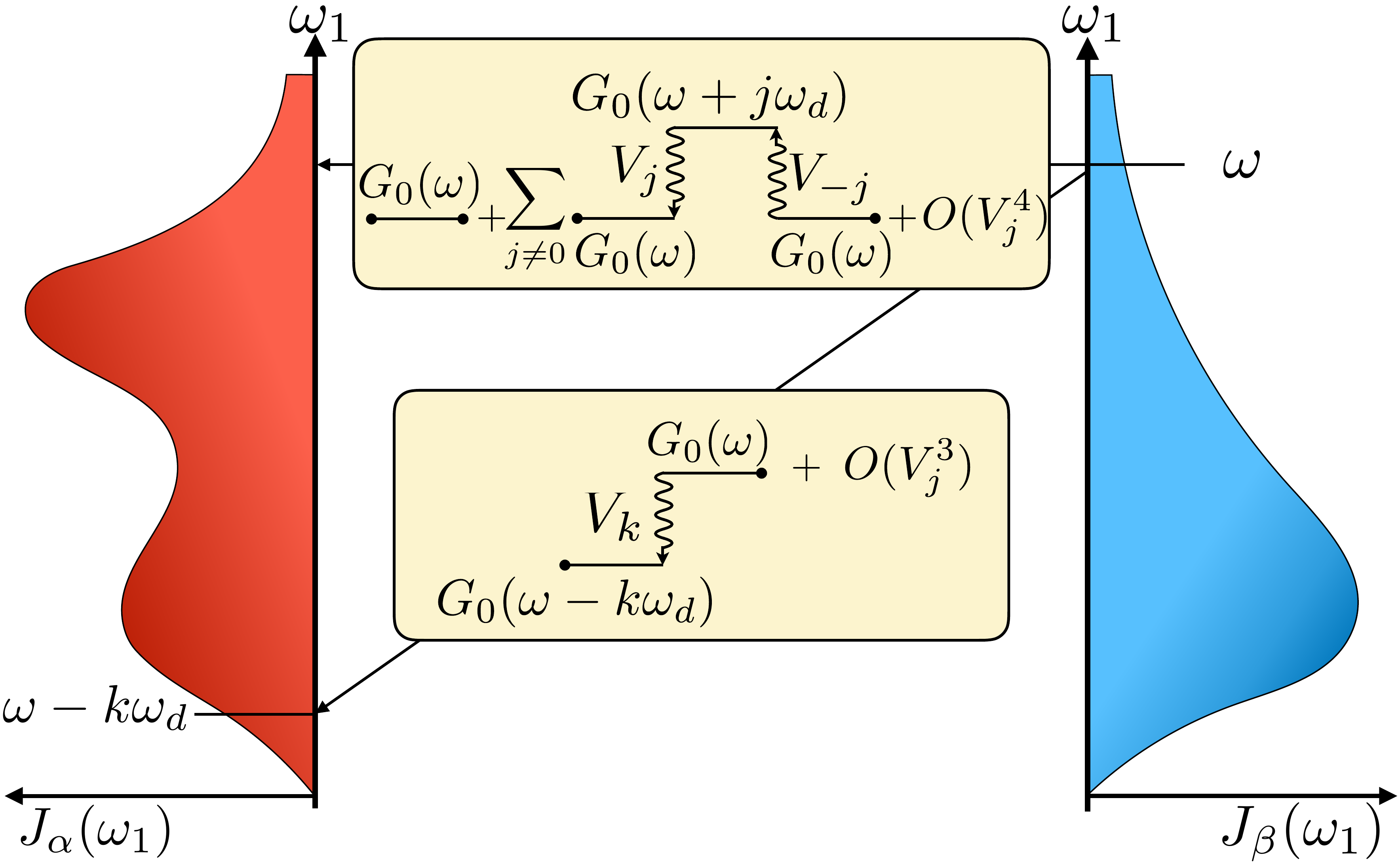}
	\caption{Diagrams representing the lowest order processes of energy exchange between reservoirs $\alpha$ and $\beta$ appearing in Eq.~\eqref{eq:A0_perturbative} and Eq.~\eqref{eq:Ak_perturbative}. Notice that there is a correspondence between the symmetry of the diagrams and the symmetry property of the matrix they represent. } 
	\label{fig:diagrams}
\end{figure}

Eq.~\eqref{eq:averaged_work} allows to define \textit{local} work rates  $\bar{\dot{W}}_\alpha$, in such a way that $ \sum_\alpha \bar{\dot{W}}_\alpha = \bar{\dot{W}}$. 
We remark that for a periodically driven network $\mathcal{T}_{\alpha\beta}(\omega)\neq \mathcal{T}_{\beta\alpha}(\omega)$, and therefore the system is no longer a symmetric heat conductor. This is the key difference between the static and driven scenarios and the cornerstone of this work. Let us explore the fundamental processes leading to such asymmetric heat exchange. For simplicity, let each reservoir be coupled to one and only one network node. Therefore, the spectral densities $J_\alpha(\omega)$ become rank-one matrices, and 
\begin{align}
r_{\alpha\beta}^k(\omega) = \pi \tilde{J}_\alpha(\omega-k\omega_d)|A_k(\omega)_{\alpha\beta}|^2\tilde{J}_\beta(\omega), \label{eq:magnitude_transition}
\end{align}
where $\tilde{J}_\alpha(\omega)$, $\tilde{J}_\beta(\omega)$ are scalar functions. The rate of a certain process depends, not only on the starting and ending reservoir modes, but also on the term $|A_k(\omega)_{\alpha\beta}|^2$---which accounts for the network conductivity for a process of exchanging $k$ energy quanta with the driving source. Self-consistently solving Eq.~\eqref{eq:eom_amplitudesk} yields
\begin{align}
A_0(\omega) &= G_0(\omega)+ \nonumber\\&\hspace{-0.3 in}+ \sum_{j\neq 0}G_0(\omega)V_j G_0(\omega+ j\omega_d) V_{-j} G_0(\omega) + O(V_j^4), \label{eq:A0_perturbative}\\
A_k(\omega) &= -G_0(\omega- k\omega_d) V_k G_0(\omega) + O(V_j^3), \hspace{0.08in} k\neq 0\label{eq:Ak_perturbative}
\end{align}
These are perturbative expansions in terms of the strength of the driving potential. Therefore, $A_k(\omega)$ can be interpreted as the fundamental processes responsible of the heat transport. Such expansions can be seen as a combination of free evolution with interactions with the periodic driving, the later causing a sudden change of the propagation frequency. This idea is depicted in the diagrams in Fig.~\ref{fig:diagrams}. Since the fundamental processes involved in the $A_0(\omega)$ are symmetric, the rate fulfills $r^0_{\alpha\beta}(\omega)=r^0_{\beta\alpha}(\omega)$. However, $A_k(\omega)$ contains asymmetric processes, and therefore, \textit{a priori} $r^k_{\alpha\beta}(\omega)\neq r^k_{\beta\alpha}(\omega)$ for $k \neq 0$. 

Notice that, particularizing this result to the case of two reservoirs can lead to a non-zero rectification coefficient, $R(\bar{\dot{Q}}_1,\bar{\dot{Q}}_1^r)$. The numerator of Eq.(\ref{eq:rectification_coef}), is given by the absolute value of
\begin{align}
\bar{\dot{Q}}_1 &+ \bar{\dot{Q}}_1^{r} = -\bar{\dot{W}}_1-\bar{\dot{W}}_1^r +\nonumber\\
&\int_\mathbb{R} \text{d}\omega \left[ \mathcal{T}_{21}(\omega)-\mathcal{T}_{12}(\omega)\right]\left(n_1(\omega) + n_2(\omega)+ 1 \right). \label{eq:num_rect_coeff}
\end{align}
Heat rectification in periodically driven systems is based on two facts: (i) the work injected into the system is a useful resource to redistribute energy and (ii) periodically driven systems exhibit new asymmetric energy exchange processes that have no analog in static systems. To better illustrate the two sources of asymmetry, we consider a network of 2 oscillators with frequencies $\omega_1$ and $\omega_2$ and equal masses ($M=\mathds{1}$), coupled to identical reservoirs ($\chi(\omega) = \tilde{\chi}(\omega)\mathds{1}$) and weakly driven by $V(t) = V_0 + 2V_1\cos(\omega_dt)$ at frequency $\omega_d$. Its expression in matrix form is
\begin{align}
V(t) = \left(\begin{matrix}
\omega_1^2 + c_0+ 2 \text{v}_1 \cos(\omega_d t) && -c_0 \\
-c_0 && \omega_2^2+c_0
\end{matrix}\right) \label{eq:V0_two},
\end{align}
where $c_0$ is the coupling constant between the oscillators, v$_1$ is the strength of the driving acting on the first oscillator. 
The normal mode frequencies of $V_0$ are $\nu_{1,2}^2 = \omega_1^2 + c_0 +\Delta/2 \mp (c_0^2 + \Delta^2/4)^{1/2}$ where $\Delta = \omega_2^2-\omega_1^2$ is the detuning between the oscillators. 
Under these conditions, the Green's function $G_0(\omega)$ is given by 
\begin{align}
\hspace{-10pt} G_0(\omega) = \left(\begin{matrix}
\frac{\cos^2\theta}{L_1(\omega)} + \frac{\sin^2\theta}{L_2(\omega)} && \frac{\sin\theta\cos\theta}{L_1(\omega)} - \frac{\sin\theta\cos\theta}{L_2(\omega)}\\
\frac{\sin\theta\cos\theta}{L_1(\omega)} - \frac{\sin\theta\cos\theta}{L_2(\omega)} &&\frac{\sin^2\theta}{L_1(\omega)} + \frac{\cos^2\theta}{L_2(\omega)} 
\end{matrix}\right), \label{eq:greens}
\end{align}
where $L_i(\omega) = \nu_i^2-\omega^2 -\text{i} \pi  \tilde{J}(\omega)$ and $\tan(2\theta) = - 2 c_0/\Delta$. For Ohmic reservoirs, the spectral density is $\tilde{J}(\omega) = \pi^{-1} 2\gamma\omega \Lambda^2 /(\omega^2+\Lambda^2)$ where  $\gamma$ is the dissipation strength, and $\Lambda$ a high-frequency cutoff. Accordingly, using fluctuation-dissipation and Kramers-Kronig relations, one obtains $\tilde{\chi}(\omega) = 2 \gamma \Lambda^2 /(\Lambda_\alpha -\text{i}\omega)$ \cite{Gonzalez2017}.

\begin{figure}[t]
	\includegraphics[width=1\linewidth]{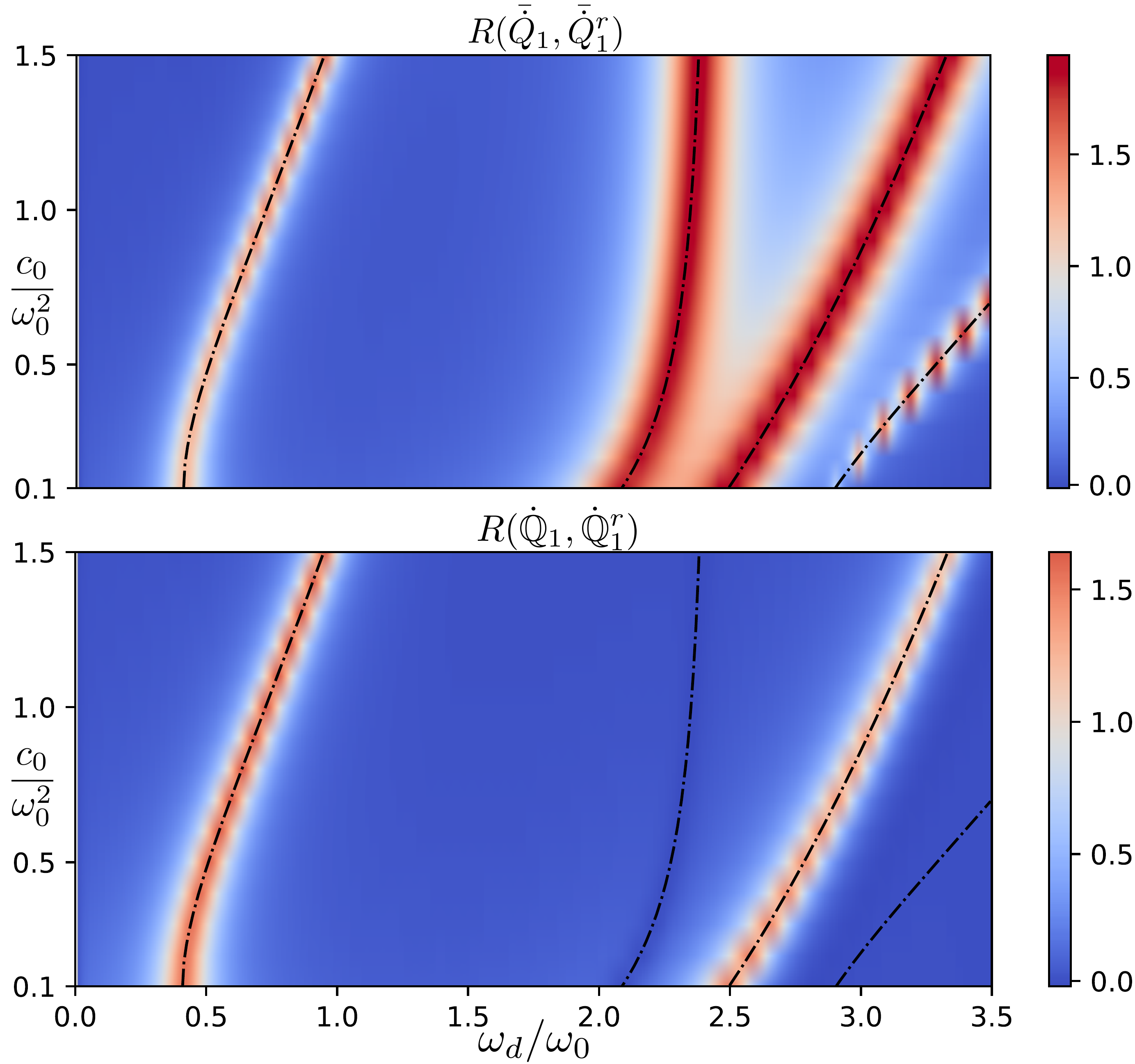}
	\caption{(Color online) Rectification coefficient as a function of the driving frequency and the coupling constant for: (top) using the full current $\bar{\dot{Q}}_1$; (bottom) using the static quasi-current $\dot{\mathbb{Q}}_1$. Black dashed lines lie at frequencies $\omega_d = \nu_i \pm \nu_j$ for $i,j = 1,2$. The parameters are set to: $\omega_1= 2\omega_0$, $\omega_2 = \omega_0$, $T_1 = 1.2 T_2 =  1.2\hbar \omega_0/k_B$, $\text{v}_1 = 0.1\omega_0^2$, $\gamma = 0.01\omega_0$, and $\Lambda = 10 \omega_0$.} 
	\label{fig:rectification}
\end{figure}

 In Fig.~\ref{fig:rectification} (top) we show the rectification coefficient versus the driving frequency $\omega_d$ and the coupling constant $c_0$. Notice that the driven network reaches $R(\bar{\dot{Q}}_1,\bar{\dot{Q}}_1^r)=1$ indicating  that the heat flux $\bar{\dot{Q}}_1$ is completely suppressed in one of the configurations. In fact, such system even attains $R(\bar{\dot{Q}}_1,\bar{\dot{Q}}_1^r)\geq1$, a regime only possible when the network conducts heat against thermal gradient in one of the configurations. The regions with non-zero rectification correspond to driving frequencies of $\omega_d = \nu_i \pm \nu_j $ for $i,j = 1,2$ such that $\omega_d\geq 0$.

To study explicitly the asymmetry of the dynamical heat transfer matrix, we define the static quasi-currents $\dot{\mathbb{Q}}_\alpha = \bar{\dot{W}}_\alpha + \bar{\dot{Q}}_\alpha$ where the contribution of the work has been explicitly singled out. Notice that they reduce to the static currents in Eq.~\eqref{eq:static_current_final} as the driving is turned off. Moreover, they fulfill the static first law $\dot{\mathbb{Q}}_1 +\dot{\mathbb{Q}}_2=0$. In this case, it is also possible to define the rectification coefficient $R(\dot{\mathbb{Q}}_1,\dot{\mathbb{Q}}_1^r)$, which is proportional to
\begin{align}
|\dot{\mathbb{Q}}_1 + \dot{\mathbb{Q}}_1^{r}| = \left|\int_\mathbb{R} \text{d}\omega \left[ \mathcal{T}_{21}(\omega)-\mathcal{T}_{12}(\omega)\right]\left(n_1(\omega) + n_2(\omega)+ 1 \right)\right|. \label{eq:qstatic_rect_coeff}
\end{align}
From Eq.~\eqref{eq:qstatic_rect_coeff}, and the same scenario with $N=2$, it follows that highly asymmetric heat transport requires that $|G_0(\omega-k\omega_d)_{21} G_0(\omega)_{11}|^2$ and $|G_0(\omega-k\omega_d)_{11} G_0(\omega)_{12}|^2$ are large compared to the static heat currents and different from each other.  Inspection of Eq. \eqref{eq:greens} indicates that only for $\omega_d = \nu_2 \pm \nu_1$ both conditions are simultaneously fulfilled. Eq.~\eqref{eq:greens} also suggests that asymmetric transport can be further enhanced in detuned systems through the parameter $\theta$. More details are given in the Supplemental Material. 

In Fig.~\ref{fig:rectification} (bottom) we show the $R(\dot{\mathbb{Q}}_1,\dot{\mathbb{Q}}_1^r)$ versus the driving frequency $\omega_d$ and the coupling constant $c_0$. In comparison with Fig.~\ref{fig:rectification} (top), we no longer observe the regions of high rectification corresponding to $\omega_d = 2\nu_1, 2\nu_2$. This implies that, even though the energy of reservoir $1$ is kept constant and energy is being injected into the system, no asymmetry of the dynamical heat transfer matrix is achieved. In accordance with the previous discussion, highly asymmetric heat transport occurs at the nearby of $\omega_d = \nu_2 \pm \nu_1$ independently of the driving strength $\text{v}_1$. 

\begin{figure}
	\includegraphics[width=1\linewidth]{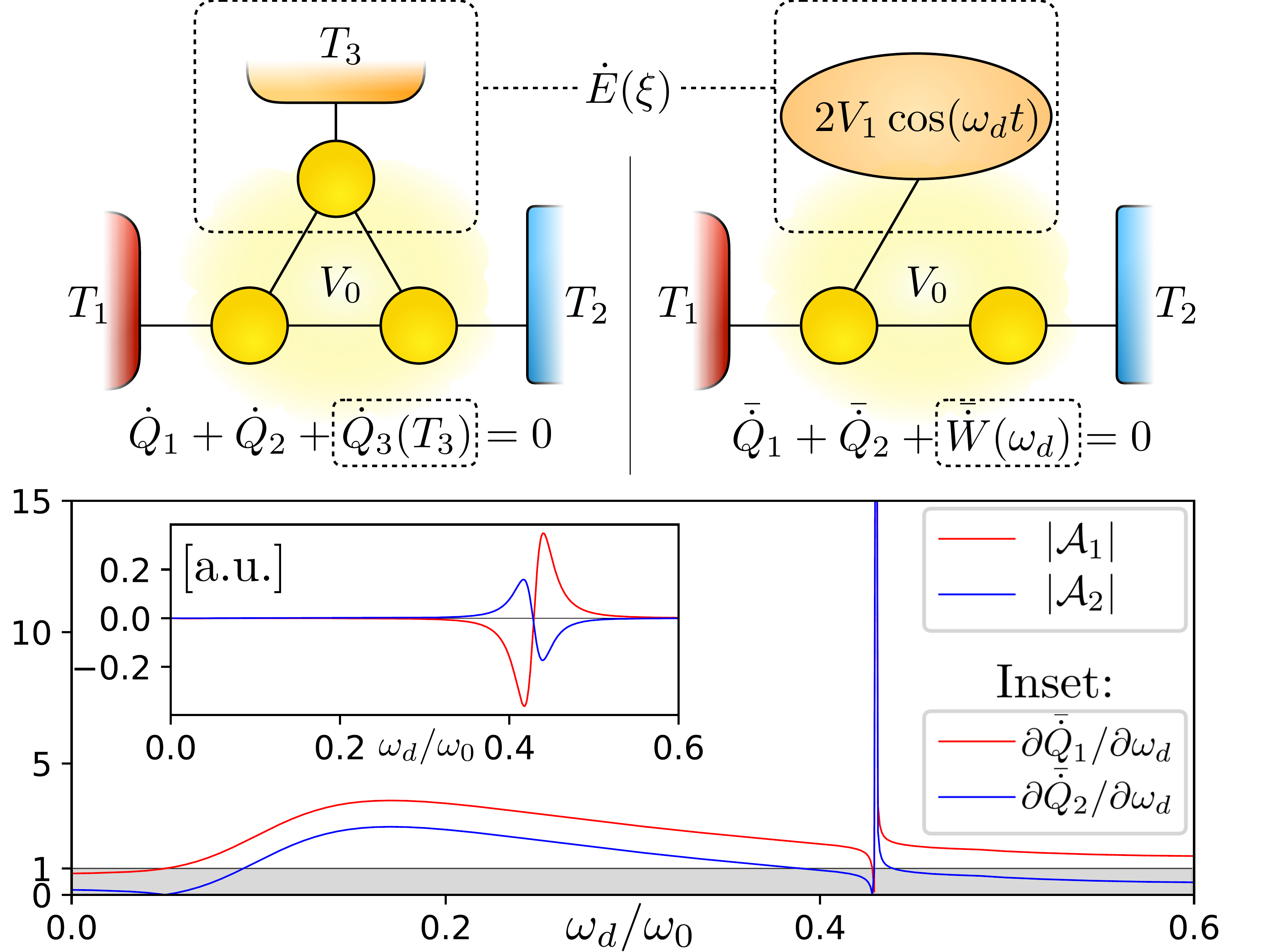}
	\caption{(Top) Scheme of a thermal bipolar transistor: (left) static with $\xi = T_3$; (right) dynamic with $\xi = \omega_d$. (Bottom) $|\mathcal{A}_{\alpha}|$ versus $\omega_d$ for $c_0 = 0.2\omega_0^2$ and same parameters as Fig.~\ref{fig:rectification}. The shadowed area represents the region without transistor effect; (inset) $\partial\bar{\dot{Q}}_{\alpha}/\partial \omega_d$ versus $\omega_d$.}
	\label{fig:transistor}
\end{figure}
Interestingly, $R(\dot{\mathbb{Q}}_1,\dot{\mathbb{Q}}_1^r)\neq 0$ is a necessary condition for the suitability of a detuned driven network to exhibit negative differential thermal resistance, a key property to build \textit{thermal transistors}~\cite{Li2006}. Thermal transistors have the capacity to manipulate heat currents flowing through the system using a third energy source that we denote by $\dot{E}$. 
The dynamical amplification factor $\mathcal{A}$ is defined as \cite{Joulain2016,Li2006}
\begin{align}
\mathcal{A}_{\alpha} \coloneqq \frac{\text{d} \dot{Q}_{\alpha}}{\text{d}\dot{E}},
\end{align}
and measures the ability of the system to \textit{amplify} a change in the third energy source. To implement this change we use a control parameter $\xi$, while keeping everything else constant. Again we compare the performance of static and driven harmonic networks. For the static case, we consider $\dot{E}(\xi)=\dot{Q}_3(T_3)$, the heat current coming from a third reservoir at temperature $T_3$, while for the driven one $\dot{E}(\xi)=\bar{\dot{W}}(\omega_d)$, the average input work of a periodic driving of frequency $\omega_d$ (see Fig.~\ref{fig:transistor}). Notice that the first law constraints the values of the dynamical amplification factors to $\mathcal{A}_1 + \mathcal{A}_2 +1 = 0$. A system is a good thermal transistor if $|\mathcal{A}_{\alpha}| \geq 1$ . From Eq.~\eqref{eq:static_current_final} follows that for the static harmonic network, $|\mathcal{A}_{\alpha}| \leq 1$ is always fulfilled. However, as shown in the bottom panel of Fig.~\ref{fig:transistor}, this is no longer true for the driven network.


In summary, using the fact that open-system dynamics can be profoundly modified by a time-periodic perturbation, we have demonstrated heat rectification in quantum linear systems; an effect strictly forbidden in the static case. We have derived analytical expressions for the steady-state heat currents and identified the fundamental processes leading to the asymmetry of the dynamical heat transfer matrix. Such asymmetry has a key role in the feasibility of periodically driven harmonic networks as, for instance, heat transistors. The formalism used here may find applications in other problems involving spatial symmetry breaking in open systems by a time-periodic perturbation. \\

\begin{acknowledgments}
	We thank N. Freitas and G. Muga for useful discussions. We acknowledge financial support from the Spanish MINECO (AEI/FEDER EU): FIS2016-80681-P, FIS2016-80773-P, FIS2015-67161-P; Severo Ochoa SEV-2015-0522. Generalitat de Catalunya: CIRIT (2017-SGR-966, 2017-SGR-1381), CERCA Program and FI-2018-B01134, Fundaci\'{o} Privada Cellex and Basque Government (Grant No. IT986-16).
\end{acknowledgments}

\bibliographystyle{apsrev4-1}
\bibliography{Refs}

\begin{thebibliography}{25}%
\makeatletter
\providecommand \@ifxundefined [1]{%
 \@ifx{#1\undefined}
}%
\providecommand \@ifnum [1]{%
 \ifnum #1\expandafter \@firstoftwo
 \else \expandafter \@secondoftwo
 \fi
}%
\providecommand \@ifx [1]{%
 \ifx #1\expandafter \@firstoftwo
 \else \expandafter \@secondoftwo
 \fi
}%
\providecommand \natexlab [1]{#1}%
\providecommand \enquote  [1]{``#1''}%
\providecommand \bibnamefont  [1]{#1}%
\providecommand \bibfnamefont [1]{#1}%
\providecommand \citenamefont [1]{#1}%
\providecommand \href@noop [0]{\@secondoftwo}%
\providecommand \href [0]{\begingroup \@sanitize@url \@href}%
\providecommand \@href[1]{\@@startlink{#1}\@@href}%
\providecommand \@@href[1]{\endgroup#1\@@endlink}%
\providecommand \@sanitize@url [0]{\catcode `\\12\catcode `\$12\catcode
  `\&12\catcode `\#12\catcode `\^12\catcode `\_12\catcode `\%12\relax}%
\providecommand \@@startlink[1]{}%
\providecommand \@@endlink[0]{}%
\providecommand \url  [0]{\begingroup\@sanitize@url \@url }%
\providecommand \@url [1]{\endgroup\@href {#1}{\urlprefix }}%
\providecommand \urlprefix  [0]{URL }%
\providecommand \Eprint [0]{\href }%
\providecommand \doibase [0]{http://dx.doi.org/}%
\providecommand \selectlanguage [0]{\@gobble}%
\providecommand \bibinfo  [0]{\@secondoftwo}%
\providecommand \bibfield  [0]{\@secondoftwo}%
\providecommand \translation [1]{[#1]}%
\providecommand \BibitemOpen [0]{}%
\providecommand \bibitemStop [0]{}%
\providecommand \bibitemNoStop [0]{.\EOS\space}%
\providecommand \EOS [0]{\spacefactor3000\relax}%
\providecommand \BibitemShut  [1]{\csname bibitem#1\endcsname}%
\let\auto@bib@innerbib\@empty
\bibitem [{\citenamefont {Li}\ \emph {et~al.}(2012)\citenamefont {Li},
  \citenamefont {Ren}, \citenamefont {Wang}, \citenamefont {Zhang},
  \citenamefont {H\"anggi},\ and\ \citenamefont {Li}}]{RevModPhys.84.1045}%
  \BibitemOpen
  \bibfield  {author} {\bibinfo {author} {\bibfnamefont {N.}~\bibnamefont
  {Li}}, \bibinfo {author} {\bibfnamefont {J.}~\bibnamefont {Ren}}, \bibinfo
  {author} {\bibfnamefont {L.}~\bibnamefont {Wang}}, \bibinfo {author}
  {\bibfnamefont {G.}~\bibnamefont {Zhang}}, \bibinfo {author} {\bibfnamefont
  {P.}~\bibnamefont {H\"anggi}}, \ and\ \bibinfo {author} {\bibfnamefont
  {B.}~\bibnamefont {Li}},\ }\href {\doibase 10.1103/RevModPhys.84.1045}
  {\bibfield  {journal} {\bibinfo  {journal} {Rev. Mod. Phys.}\ }\textbf
  {\bibinfo {volume} {84}},\ \bibinfo {pages} {1045} (\bibinfo {year}
  {2012})}\BibitemShut {NoStop}%
\bibitem [{\citenamefont {Chang}\ \emph {et~al.}(2006)\citenamefont {Chang},
  \citenamefont {Okawa}, \citenamefont {Majumdar},\ and\ \citenamefont
  {Zettl}}]{Chang1121}%
  \BibitemOpen
  \bibfield  {author} {\bibinfo {author} {\bibfnamefont {C.~W.}\ \bibnamefont
  {Chang}}, \bibinfo {author} {\bibfnamefont {D.}~\bibnamefont {Okawa}},
  \bibinfo {author} {\bibfnamefont {A.}~\bibnamefont {Majumdar}}, \ and\
  \bibinfo {author} {\bibfnamefont {A.}~\bibnamefont {Zettl}},\ }\href
  {\doibase 10.1126/science.1132898} {\bibfield  {journal} {\bibinfo  {journal}
  {Science}\ }\textbf {\bibinfo {volume} {314}},\ \bibinfo {pages} {1121}
  (\bibinfo {year} {2006})},\ \Eprint
  {http://arxiv.org/abs/http://science.sciencemag.org/content/314/5802/1121.full.pdf}
  {http://science.sciencemag.org/content/314/5802/1121.full.pdf} \BibitemShut
  {NoStop}%
\bibitem [{\citenamefont {Benenti}\ \emph {et~al.}(2016)\citenamefont
  {Benenti}, \citenamefont {Casati}, \citenamefont {Mej{\'i}a-Monasterio},\
  and\ \citenamefont {Peyrard}}]{Benenti2016}%
  \BibitemOpen
  \bibfield  {author} {\bibinfo {author} {\bibfnamefont {G.}~\bibnamefont
  {Benenti}}, \bibinfo {author} {\bibfnamefont {G.}~\bibnamefont {Casati}},
  \bibinfo {author} {\bibfnamefont {C.}~\bibnamefont {Mej{\'i}a-Monasterio}}, \
  and\ \bibinfo {author} {\bibfnamefont {M.}~\bibnamefont {Peyrard}},\
  }\enquote {\bibinfo {title} {From thermal rectifiers to thermoelectric
  devices},}\ \ (\bibinfo  {publisher} {Springer International Publishing},\
  \bibinfo {address} {Cham},\ \bibinfo {year} {2016})\ pp.\ \bibinfo {pages}
  {365--407}\BibitemShut {NoStop}%
\bibitem [{\citenamefont {Terraneo}\ \emph {et~al.}(2002)\citenamefont
  {Terraneo}, \citenamefont {Peyrard},\ and\ \citenamefont
  {Casati}}]{PhysRevLett.88.094302}%
  \BibitemOpen
  \bibfield  {author} {\bibinfo {author} {\bibfnamefont {M.}~\bibnamefont
  {Terraneo}}, \bibinfo {author} {\bibfnamefont {M.}~\bibnamefont {Peyrard}}, \
  and\ \bibinfo {author} {\bibfnamefont {G.}~\bibnamefont {Casati}},\ }\href
  {\doibase 10.1103/PhysRevLett.88.094302} {\bibfield  {journal} {\bibinfo
  {journal} {Phys. Rev. Lett.}\ }\textbf {\bibinfo {volume} {88}},\ \bibinfo
  {pages} {094302} (\bibinfo {year} {2002})}\BibitemShut {NoStop}%
\bibitem [{\citenamefont {van Zwol}\ \emph {et~al.}(2012)\citenamefont {van
  Zwol}, \citenamefont {Ranno},\ and\ \citenamefont
  {Chevrier}}]{PhysRevLett.108.234301}%
  \BibitemOpen
  \bibfield  {author} {\bibinfo {author} {\bibfnamefont {P.~J.}\ \bibnamefont
  {van Zwol}}, \bibinfo {author} {\bibfnamefont {L.}~\bibnamefont {Ranno}}, \
  and\ \bibinfo {author} {\bibfnamefont {J.}~\bibnamefont {Chevrier}},\ }\href
  {\doibase 10.1103/PhysRevLett.108.234301} {\bibfield  {journal} {\bibinfo
  {journal} {Phys. Rev. Lett.}\ }\textbf {\bibinfo {volume} {108}},\ \bibinfo
  {pages} {234301} (\bibinfo {year} {2012})}\BibitemShut {NoStop}%
\bibitem [{\citenamefont {{Peyrard, M.}}(2006)}]{refId0}%
  \BibitemOpen
  \bibfield  {author} {\bibinfo {author} {\bibnamefont {{Peyrard, M.}}},\
  }\href {\doibase 10.1209/epl/i2006-10223-5} {\bibfield  {journal} {\bibinfo
  {journal} {Europhys. Lett.}\ }\textbf {\bibinfo {volume} {76}},\ \bibinfo
  {pages} {49} (\bibinfo {year} {2006})}\BibitemShut {NoStop}%
\bibitem [{\citenamefont {Liu}\ \emph {et~al.}(2014)\citenamefont {Liu},
  \citenamefont {Zhou}, \citenamefont {Tang},\ and\ \citenamefont
  {Chen}}]{liu2014important}%
  \BibitemOpen
  \bibfield  {author} {\bibinfo {author} {\bibfnamefont {Y.-Y.}\ \bibnamefont
  {Liu}}, \bibinfo {author} {\bibfnamefont {W.-X.}\ \bibnamefont {Zhou}},
  \bibinfo {author} {\bibfnamefont {L.-M.}\ \bibnamefont {Tang}}, \ and\
  \bibinfo {author} {\bibfnamefont {K.-Q.}\ \bibnamefont {Chen}},\ }\href@noop
  {} {\bibfield  {journal} {\bibinfo  {journal} {Applied Physics Letters}\
  }\textbf {\bibinfo {volume} {105}},\ \bibinfo {pages} {203111} (\bibinfo
  {year} {2014})}\BibitemShut {NoStop}%
\bibitem [{\citenamefont {{Pons, M.}}\ \emph {et~al.}(2017)\citenamefont
  {{Pons, M.}}, \citenamefont {{Cui, Y. Y.}}, \citenamefont {{Ruschhaupt, A.}},
  \citenamefont {{Sim\'on, M. A.}},\ and\ \citenamefont {{Muga, J.
  G.}}}]{refId009}%
  \BibitemOpen
  \bibfield  {author} {\bibinfo {author} {\bibnamefont {{Pons, M.}}}, \bibinfo
  {author} {\bibnamefont {{Cui, Y. Y.}}}, \bibinfo {author} {\bibnamefont
  {{Ruschhaupt, A.}}}, \bibinfo {author} {\bibnamefont {{Sim\'on, M. A.}}}, \
  and\ \bibinfo {author} {\bibnamefont {{Muga, J. G.}}},\ }\href {\doibase
  10.1209/0295-5075/119/64001} {\bibfield  {journal} {\bibinfo  {journal}
  {EPL}\ }\textbf {\bibinfo {volume} {119}},\ \bibinfo {pages} {64001}
  (\bibinfo {year} {2017})}\BibitemShut {NoStop}%
\bibitem [{\citenamefont {Arrachea}\ \emph {et~al.}(2012)\citenamefont
  {Arrachea}, \citenamefont {Mucciolo}, \citenamefont {Chamon},\ and\
  \citenamefont {Capaz}}]{Arrachea2012}%
  \BibitemOpen
  \bibfield  {author} {\bibinfo {author} {\bibfnamefont {L.}~\bibnamefont
  {Arrachea}}, \bibinfo {author} {\bibfnamefont {E.~R.}\ \bibnamefont
  {Mucciolo}}, \bibinfo {author} {\bibfnamefont {C.}~\bibnamefont {Chamon}}, \
  and\ \bibinfo {author} {\bibfnamefont {R.~B.}\ \bibnamefont {Capaz}},\ }\href
  {\doibase 10.1103/PhysRevB.86.125424} {\bibfield  {journal} {\bibinfo
  {journal} {Physical Review B - Condensed Matter and Materials Physics}\ }
  (\bibinfo {year} {2012}),\ 10.1103/PhysRevB.86.125424},\ \Eprint
  {http://arxiv.org/abs/arXiv:1203.2561v2} {arXiv:arXiv:1203.2561v2}
  \BibitemShut {NoStop}%
\bibitem [{\citenamefont {Freitas}\ and\ \citenamefont
  {Paz}(2017)}]{Freitas2017}%
  \BibitemOpen
  \bibfield  {author} {\bibinfo {author} {\bibfnamefont {N.}~\bibnamefont
  {Freitas}}\ and\ \bibinfo {author} {\bibfnamefont {J.~P.}\ \bibnamefont
  {Paz}},\ }\href {\doibase 10.1103/PhysRevE.95.012146} {\bibfield  {journal}
  {\bibinfo  {journal} {Physical Review E}\ }\textbf {\bibinfo {volume} {95}}
  (\bibinfo {year} {2017}),\ 10.1103/PhysRevE.95.012146},\ \Eprint
  {http://arxiv.org/abs/1607.04234} {arXiv:1607.04234} \BibitemShut {NoStop}%
\bibitem [{\citenamefont {Scheibner}\ \emph {et~al.}(2008)\citenamefont
  {Scheibner}, \citenamefont {König}, \citenamefont {Reuter}, \citenamefont
  {Wieck}, \citenamefont {Gould}, \citenamefont {Buhmann},\ and\ \citenamefont
  {Molenkamp}}]{1367-2630-10-8-083016}%
  \BibitemOpen
  \bibfield  {author} {\bibinfo {author} {\bibfnamefont {R.}~\bibnamefont
  {Scheibner}}, \bibinfo {author} {\bibfnamefont {M.}~\bibnamefont {König}},
  \bibinfo {author} {\bibfnamefont {D.}~\bibnamefont {Reuter}}, \bibinfo
  {author} {\bibfnamefont {A.~D.}\ \bibnamefont {Wieck}}, \bibinfo {author}
  {\bibfnamefont {C.}~\bibnamefont {Gould}}, \bibinfo {author} {\bibfnamefont
  {H.}~\bibnamefont {Buhmann}}, \ and\ \bibinfo {author} {\bibfnamefont
  {L.~W.}\ \bibnamefont {Molenkamp}},\ }\href
  {http://stacks.iop.org/1367-2630/10/i=8/a=083016} {\bibfield  {journal}
  {\bibinfo  {journal} {New Journal of Physics}\ }\textbf {\bibinfo {volume}
  {10}},\ \bibinfo {pages} {083016} (\bibinfo {year} {2008})}\BibitemShut
  {NoStop}%
\bibitem [{\citenamefont {Ruokola}\ \emph {et~al.}(2009)\citenamefont
  {Ruokola}, \citenamefont {Ojanen},\ and\ \citenamefont
  {Jauho}}]{PhysRevB.79.144306}%
  \BibitemOpen
  \bibfield  {author} {\bibinfo {author} {\bibfnamefont {T.}~\bibnamefont
  {Ruokola}}, \bibinfo {author} {\bibfnamefont {T.}~\bibnamefont {Ojanen}}, \
  and\ \bibinfo {author} {\bibfnamefont {A.-P.}\ \bibnamefont {Jauho}},\ }\href
  {\doibase 10.1103/PhysRevB.79.144306} {\bibfield  {journal} {\bibinfo
  {journal} {Phys. Rev. B}\ }\textbf {\bibinfo {volume} {79}},\ \bibinfo
  {pages} {144306} (\bibinfo {year} {2009})}\BibitemShut {NoStop}%
\bibitem [{\citenamefont {Joulain}\ \emph {et~al.}(2016)\citenamefont
  {Joulain}, \citenamefont {Drevillon}, \citenamefont {Ezzahri},\ and\
  \citenamefont {Ordonez-Miranda}}]{Joulain2016}%
  \BibitemOpen
  \bibfield  {author} {\bibinfo {author} {\bibfnamefont {K.}~\bibnamefont
  {Joulain}}, \bibinfo {author} {\bibfnamefont {J.}~\bibnamefont {Drevillon}},
  \bibinfo {author} {\bibfnamefont {Y.}~\bibnamefont {Ezzahri}}, \ and\
  \bibinfo {author} {\bibfnamefont {J.}~\bibnamefont {Ordonez-Miranda}},\
  }\href {\doibase 10.1103/PhysRevLett.116.200601} {\bibfield  {journal}
  {\bibinfo  {journal} {Physical Review Letters}\ }\textbf {\bibinfo {volume}
  {116}} (\bibinfo {year} {2016}),\ 10.1103/PhysRevLett.116.200601},\ \Eprint
  {http://arxiv.org/abs/1602.04175} {arXiv:1602.04175} \BibitemShut {NoStop}%
\bibitem [{\citenamefont {{Wang}}\ \emph {et~al.}(2018)\citenamefont {{Wang}},
  \citenamefont {{Xu}}, \citenamefont {{Liu}},\ and\ \citenamefont
  {{Gao}}}]{2018arXiv180604794W}%
  \BibitemOpen
  \bibfield  {author} {\bibinfo {author} {\bibfnamefont {C.}~\bibnamefont
  {{Wang}}}, \bibinfo {author} {\bibfnamefont {D.}~\bibnamefont {{Xu}}},
  \bibinfo {author} {\bibfnamefont {H.}~\bibnamefont {{Liu}}}, \ and\ \bibinfo
  {author} {\bibfnamefont {X.}~\bibnamefont {{Gao}}},\ }\href@noop {}
  {\bibfield  {journal} {\bibinfo  {journal} {arXiv e-prints}\ } (\bibinfo
  {year} {2018})},\ \Eprint {http://arxiv.org/abs/1806.04794} {arXiv:1806.04794
  [quant-ph]} \BibitemShut {NoStop}%
\bibitem [{\citenamefont {Wu}\ \emph {et~al.}(2009)\citenamefont {Wu},
  \citenamefont {Yu},\ and\ \citenamefont {Segal}}]{PhysRevE.80.041103}%
  \BibitemOpen
  \bibfield  {author} {\bibinfo {author} {\bibfnamefont {L.-A.}\ \bibnamefont
  {Wu}}, \bibinfo {author} {\bibfnamefont {C.~X.}\ \bibnamefont {Yu}}, \ and\
  \bibinfo {author} {\bibfnamefont {D.}~\bibnamefont {Segal}},\ }\href
  {\doibase 10.1103/PhysRevE.80.041103} {\bibfield  {journal} {\bibinfo
  {journal} {Phys. Rev. E}\ }\textbf {\bibinfo {volume} {80}},\ \bibinfo
  {pages} {041103} (\bibinfo {year} {2009})}\BibitemShut {NoStop}%
\bibitem [{\citenamefont {Roberts}\ and\ \citenamefont
  {Walker}(2011)}]{Roberts2011}%
  \BibitemOpen
  \bibfield  {author} {\bibinfo {author} {\bibfnamefont {N.~A.}\ \bibnamefont
  {Roberts}}\ and\ \bibinfo {author} {\bibfnamefont {D.~G.}\ \bibnamefont
  {Walker}},\ }\href {\doibase 10.1016/j.ijthermalsci.2010.12.004} {\bibfield
  {journal} {\bibinfo  {journal} {International Journal of Thermal Sciences}\ }
  (\bibinfo {year} {2011}),\ 10.1016/j.ijthermalsci.2010.12.004}\BibitemShut
  {NoStop}%
\bibitem [{\citenamefont {Grabert}\ \emph {et~al.}(1984)\citenamefont
  {Grabert}, \citenamefont {Weiss},\ and\ \citenamefont
  {Talkner}}]{Grabert1984}%
  \BibitemOpen
  \bibfield  {author} {\bibinfo {author} {\bibfnamefont {H.}~\bibnamefont
  {Grabert}}, \bibinfo {author} {\bibfnamefont {U.}~\bibnamefont {Weiss}}, \
  and\ \bibinfo {author} {\bibfnamefont {P.}~\bibnamefont {Talkner}},\ }\href
  {\doibase 10.1007/BF01307505} {\bibfield  {journal} {\bibinfo  {journal}
  {Zeitschrift f{\"{u}}r Physik B Condensed Matter}\ }\textbf {\bibinfo
  {volume} {55}},\ \bibinfo {pages} {87} (\bibinfo {year} {1984})}\BibitemShut
  {NoStop}%
\bibitem [{\citenamefont {Breuer}\ and\ \citenamefont
  {Petruccione}(2007)}]{Breuer2007}%
  \BibitemOpen
  \bibfield  {author} {\bibinfo {author} {\bibfnamefont {H.-P.}\ \bibnamefont
  {Breuer}}\ and\ \bibinfo {author} {\bibfnamefont {F.}~\bibnamefont
  {Petruccione}},\ }\href {\doibase 10.1093/acprof:oso/9780199213900.001.0001}
  {\emph {\bibinfo {title} {The Theory of Open Quantum Systems}}}\ (\bibinfo
  {year} {2007})\ pp.\ \bibinfo {pages} {1--656}\BibitemShut {NoStop}%
\bibitem [{\citenamefont {Kubo}(1966)}]{Kubo1966}%
  \BibitemOpen
  \bibfield  {author} {\bibinfo {author} {\bibfnamefont {R.}~\bibnamefont
  {Kubo}},\ }\href {\doibase 10.1088/0034-4885/29/1/306} {\bibfield  {journal}
  {\bibinfo  {journal} {Reports on Progress in Physics}\ }\textbf {\bibinfo
  {volume} {29}},\ \bibinfo {pages} {255} (\bibinfo {year} {1966})}\BibitemShut
  {NoStop}%
\bibitem [{\citenamefont {Esposito}\ \emph {et~al.}(2010)\citenamefont
  {Esposito}, \citenamefont {Lindenberg},\ and\ \citenamefont {{Van Den
  Broeck}}}]{Esposito2010}%
  \BibitemOpen
  \bibfield  {author} {\bibinfo {author} {\bibfnamefont {M.}~\bibnamefont
  {Esposito}}, \bibinfo {author} {\bibfnamefont {K.}~\bibnamefont
  {Lindenberg}}, \ and\ \bibinfo {author} {\bibfnamefont {C.}~\bibnamefont
  {{Van Den Broeck}}},\ }\href {\doibase 10.1088/1367-2630/12/1/013013}
  {\bibfield  {journal} {\bibinfo  {journal} {New Journal of Physics}\ }
  (\bibinfo {year} {2010}),\ 10.1088/1367-2630/12/1/013013},\ \Eprint
  {http://arxiv.org/abs/0908.1125} {arXiv:0908.1125} \BibitemShut {NoStop}%
\bibitem [{\citenamefont {Dhar}\ and\ \citenamefont {Roy}(2006)}]{Dhar2006}%
  \BibitemOpen
  \bibfield  {author} {\bibinfo {author} {\bibfnamefont {A.}~\bibnamefont
  {Dhar}}\ and\ \bibinfo {author} {\bibfnamefont {D.}~\bibnamefont {Roy}},\
  }\href {\doibase 10.1007/s10955-006-9235-3} {\bibfield  {journal} {\bibinfo
  {journal} {Journal of Statistical Physics}\ }\textbf {\bibinfo {volume}
  {125}},\ \bibinfo {pages} {805} (\bibinfo {year} {2006})},\ \Eprint
  {http://arxiv.org/abs/0606465} {arXiv:0606465 [cond-mat]} \BibitemShut
  {NoStop}%
\bibitem [{\citenamefont {Alicki}\ \emph {et~al.}(2006)\citenamefont {Alicki},
  \citenamefont {Lidar},\ and\ \citenamefont {Zanardi}}]{PhysRevA.73.052311}%
  \BibitemOpen
  \bibfield  {author} {\bibinfo {author} {\bibfnamefont {R.}~\bibnamefont
  {Alicki}}, \bibinfo {author} {\bibfnamefont {D.~A.}\ \bibnamefont {Lidar}}, \
  and\ \bibinfo {author} {\bibfnamefont {P.}~\bibnamefont {Zanardi}},\ }\href
  {\doibase 10.1103/PhysRevA.73.052311} {\bibfield  {journal} {\bibinfo
  {journal} {Phys. Rev. A}\ }\textbf {\bibinfo {volume} {73}},\ \bibinfo
  {pages} {052311} (\bibinfo {year} {2006})}\BibitemShut {NoStop}%
\bibitem [{\citenamefont {Freitas}\ and\ \citenamefont
  {Paz}(2018)}]{Freitas2018}%
  \BibitemOpen
  \bibfield  {author} {\bibinfo {author} {\bibfnamefont {N.}~\bibnamefont
  {Freitas}}\ and\ \bibinfo {author} {\bibfnamefont {J.~P.}\ \bibnamefont
  {Paz}},\ }\href {\doibase 10.1103/PhysRevA.97.032104} {\bibfield  {journal}
  {\bibinfo  {journal} {Physical Review A}\ }\textbf {\bibinfo {volume} {97}}
  (\bibinfo {year} {2018}),\ 10.1103/PhysRevA.97.032104},\ \Eprint
  {http://arxiv.org/abs/1710.11554} {arXiv:1710.11554} \BibitemShut {NoStop}%
\bibitem [{\citenamefont {Gonz{\'{a}}lez}\ \emph {et~al.}(2017)\citenamefont
  {Gonz{\'{a}}lez}, \citenamefont {Correa}, \citenamefont {Nocerino},
  \citenamefont {Palao}, \citenamefont {Alonso},\ and\ \citenamefont
  {Adesso}}]{Gonzalez2017}%
  \BibitemOpen
  \bibfield  {author} {\bibinfo {author} {\bibfnamefont {J.~O.}\ \bibnamefont
  {Gonz{\'{a}}lez}}, \bibinfo {author} {\bibfnamefont {L.~A.}\ \bibnamefont
  {Correa}}, \bibinfo {author} {\bibfnamefont {G.}~\bibnamefont {Nocerino}},
  \bibinfo {author} {\bibfnamefont {J.~P.}\ \bibnamefont {Palao}}, \bibinfo
  {author} {\bibfnamefont {D.}~\bibnamefont {Alonso}}, \ and\ \bibinfo {author}
  {\bibfnamefont {G.}~\bibnamefont {Adesso}},\ }\href {\doibase
  10.1142/S1230161217400108} {\bibfield  {journal} {\bibinfo  {journal} {Open
  Systems {\&} Information Dynamics}\ }\textbf {\bibinfo {volume} {24}},\
  \bibinfo {pages} {1740010} (\bibinfo {year} {2017})}\BibitemShut {NoStop}%
\bibitem [{\citenamefont {Li}\ \emph {et~al.}(2006)\citenamefont {Li},
  \citenamefont {Wang},\ and\ \citenamefont {Casati}}]{Li2006}%
  \BibitemOpen
  \bibfield  {author} {\bibinfo {author} {\bibfnamefont {B.}~\bibnamefont
  {Li}}, \bibinfo {author} {\bibfnamefont {L.}~\bibnamefont {Wang}}, \ and\
  \bibinfo {author} {\bibfnamefont {G.}~\bibnamefont {Casati}},\ }\href
  {\doibase 10.1063/1.2191730} {\bibfield  {journal} {\bibinfo  {journal}
  {Applied Physics Letters}\ }\textbf {\bibinfo {volume} {88}} (\bibinfo {year}
  {2006}),\ 10.1063/1.2191730},\ \Eprint {http://arxiv.org/abs/0410172}
  {arXiv:0410172 [cond-mat]} \BibitemShut {NoStop}%
\end{thebibliography}%
\newpage
\onecolumngrid
\appendix
%
\newpage
\section{Quantum Langevin Equation}
\label{sec:App1}
The Heisenberg equation of motion for any observable $O$ reads as 
\begin{align}\label{eq-Heis}
\dot{O} = \text{i}/\hbar [H,O]+\partial_t O,
\end{align}
with $H= H_S + \sum_\alpha H_{SR_\alpha} + \sum_\alpha H_{R_\alpha}$ being the total Hamiltonian.
By means of the bosonic algebra, and by using Eqs.~(2-4) of the main text, one finds the equations of motion for the system and reservoir degrees of freedom:
\begin{align}
&\dot{X}_\alpha = M_\alpha^{-1} P_\alpha,\label{eq:eom_xalpha}\\
&\dot{P}_\alpha = -M_\alpha \Omega_\alpha^2 X_\alpha +  C_\alpha^T X, \label{eq:eom_palpha}\\
&\dot{X} = M^{-1} P,\label{eq:eom_x}\\
&\dot{P} = -V(t) X +  \sum_\alpha C_\alpha X_\alpha.\label{eq:eom_p} 
\end{align}
By taking the time derivative of Eq.~\eqref{eq:eom_xalpha}, and combining with Eq.~\eqref{eq:eom_palpha}, we end up with a second order differential equation for the position quadrature operator of the bath $\alpha$. Under the initial condition $\rho(t_0) = \rho_S(t_0) \bigotimes_\alpha \rho_{R_\alpha}(t_0)$, the solution of this equation yields: 
\begin{align}
X_\alpha(t) =  \cos\left( \Omega_\alpha (t-t_0)\right)X_\alpha(t_0) +  (M_\alpha \Omega_\alpha)^{-1}\sin \left(\Omega_\alpha (t-t_0)\right)P_\alpha(t_0) + \int_{t_0}^{t} \text{d}t' (M_\alpha \Omega_\alpha)^{-1} \sin\left(\Omega_\alpha(t-t')\right)C^T_\alpha X(t'),\label{eq:solution_xalpha} 
\end{align}
where $X_\alpha(t_0)$ and $P_\alpha(t_0)$ are the initial conditions. 
In the same manner, by taking the time derivative of Eq.~(\ref{eq:eom_x}), and by substituting Eqs.~\eqref{eq:eom_p} and \eqref{eq:solution_xalpha} into it we arrive at:
\begin{align}
M\ddot{X}+V(t)X-\sum_\alpha\int_{t_0}^t\text{d}t'  C_\alpha (M_\alpha \Omega_\alpha)^{-1} \sin(\Omega_\alpha (t-t') C^T_\alpha X(t')) = \sum_\alpha C_\alpha\left(\cos\left( \Omega_\alpha (t-t_0)\right)X_\alpha(t_0) +  (M_\alpha \Omega_\alpha)^{-1}\sin \left(\Omega_\alpha (t-t_0)\right)P_\alpha(t_0)\right)
\end{align}
Since we are only interested in the steady-state behavior, 
we take the limit $t_0 \to -\infty$. Thus, we find the quantum Langevin equation:
\begin{align}
M\ddot{X} + V(t)X - \chi(t) \star X = B(t),\label{eq:the_qle}
\end{align}
with the susceptibility matrix $\chi(t)$ and the noise vector $B(t)$ given by Eq.~(6) and Eq.~(7) of the main text, respectively.

\section{Fluctuation-Dissipation Relation}
In the main text we proved a version of the fluctuation-dissipation theorem, namely $\text{Im}\chi(\omega) = \pi J(\omega)$, that is state independent. In our analysis, however, we are dealing with thermal reservoirs. In particular, in many of our calculations, we need to find the bath correlation functions at thermal equilibrium.  
Thus, here, we present a proof of the {\it thermal} version of fluctuation-dissipation theorem, that enables us to calculate the bath correlation functions. To this aim, let the thermal equilibrium density matrix of the reservoirs be $\rho_{R_\alpha}(t_0) = \exp(-H_{R_\alpha}/k_B T_\alpha)/\text{tr}[\exp(-H_{R_\alpha}/k_B T_\alpha)]$.
For such thermal state, all first order moments vanish. Moreover, the second order quadrature moments read as:
\begin{align}
&\left\langle X_\alpha(t_0)X^T_\alpha(t_0)\right\rangle = (M_\alpha\Omega_\alpha)^{-1}\left\langle P_\alpha(t_0)P^T_\alpha(t_0)\right\rangle (M_\alpha\Omega_\alpha)^{-1} = \frac{\hbar}{2}(M_\alpha\Omega_\alpha)^{-1}\coth\left(\frac{\hbar\Omega_\alpha}{2 k_B T_\alpha}\right)\label{eq:exp_diag}\\
&\left\langle X_\alpha(t_0) P^T_\alpha(t_0)\right\rangle = \left\langle P_\alpha(t_0) X^T_\alpha(t_0)\right\rangle^* =\frac{{\rm i}\hbar}{2}. \label{eq:exp_offdiag}
\end{align}
Looking back to the definition of the noise vector Eq.~(7) of the main text, it follows that
\begin{align}
\left\langle B(t) B^T(t')  \right\rangle =\frac{\hbar}{2} \sum_\alpha C_\alpha(M_\alpha\Omega_\alpha)^{-1}\left(\cos(\Omega_\alpha (t-t'))\coth\left(\frac{\hbar\Omega_\alpha}{2 k_B T_\alpha}\right)+\text{i}\sin(\Omega_\alpha(t-t')\right) C_\alpha^T.
\end{align}
By introducing the spectral density to the last expression, it can be written as
\begin{align}
\left\langle B(t) B^T(t')  \right\rangle &=\hbar \sum_\alpha \int_0^\infty \text{d}\omega J_\alpha(\omega)\cos(\omega (t-t')) (2 n_\alpha(\omega)+1) +\text{i} \hbar \sum_\alpha \int_0^\infty \text{d}\omega J_\alpha(\omega) \sin(\omega(t-t')) \nonumber\\
&=\frac{\hbar}{2}\left[ \nu(t-t') + \text{i} \eta(t-t')\right], \label{eq:bath_correlation}
\end{align}
where $\nu(t-t')$ and $\eta(t-t')$ are, respectively, the noise and damping kernels. In the spectral domain---that is, by taking the Fourier transform $f(t)\mapsto f(\omega) = \int_{-\infty}^{\infty}dt~f(t)~{\rm e}^{{\rm i}\omega t} $ of both sides---the above equation reads
\begin{align}\label{eq:bath_correlation_freq}
\left\langle B(\omega) B^T(\omega') \right\rangle =\hbar\pi\delta(\omega+\omega') \left[\nu(\omega)+\text{i} \eta(\omega) \right],
\end{align}
with $\nu(\omega) =2\pi \sum_\alpha J_\alpha(\omega)(2n_\alpha(\omega) +1)$ and $\eta(\omega) = 2\pi \sum_\alpha J_\alpha(\omega)$. 
By substituting in Eq.~\eqref{eq:bath_correlation_freq}, we obtain the desired fluctuation-dissipation relation
\begin{align}
\left\langle B(\omega) B^T(\omega') \right\rangle =  \delta(\omega+\omega') (2\pi)^2  \hbar \sum_\alpha J_\alpha(\omega) (n_\alpha(\omega) +1). \label{eq:fluc_diss}
\end{align}

\section{Thermodynamic Quantities and the Covariance Matrix}
Suppose that the steady state covariance matrix is given, the question is, how $\dot{W}$ and $\dot{Q}_\alpha$ can be calculated. 
Let us represent the covariance matrix in four blocks
\begin{align}
\Sigma = \left(\begin{matrix}
\sigma_{XX} && \sigma_{XP}\\
\sigma_{PX} && \sigma_{PP}
\end{matrix}\right),
\end{align}
with, $\sigma_{XX} = \langle XX^T \rangle$, $\sigma_{XP} = \langle XP^T + (PX^T)^T\rangle/2$ and $\sigma_{PP} = \langle PP^T \rangle$. 
Furthermore, by using the definition of local heat currents that we introduced in the main text---i.e., $\dot{Q}_\alpha \coloneqq\text{i}/\hbar \langle [H_{SR_\alpha}, H_S] \rangle$,
one can verify that 
\begin{align}
\dot{Q}_\alpha = \frac{1}{2}\left(\langle P^T M^{-1} C_\alpha X_\alpha \rangle + \langle X_{\alpha}^T C_{\alpha}^T M^{-1} P \rangle\right). \label{eq:qalpha_bath}
\end{align}
Recall our assumption that each oscillator is at most coupled to one reservoir. We denote by $\Pi_{\alpha}$ the projector onto the space of the oscillator that is coupled to the bath $\alpha$.
Thus, from Eq.~\eqref{eq:qalpha_bath} and using Eq.~\eqref{eq:eom_p} it follows that 
\begin{align}
&\dot{Q}_\alpha = \frac{1}{2}\tr\left[\Pi_\alpha \frac{\text{d}}{\text{d}t}\sigma_{PP}M^{-1}\right] + \tr\left[ \Pi_\alpha V(t) \sigma_{XP} M^{-1}\right], \label{eq:current_cov}
\end{align}
where we have used the property of the bilinear product $\left< \Psi_1^T \tilde{M}  \Psi_2\right> = \tr\left[ \tilde{M}^T \left<\Psi_1 \Psi_2^T\right> \right]$ where $\Psi_1$ and $\Psi_1$ are vector operators and $\tilde{M}$ a scalar matrix. The work rate is easily calculated through $\dot{W} = \tr\left[\frac{\partial}{\partial t} V(t) \sigma_{XX}\right]$.
\subsection{Static Heat Currents}
For a static network $V(t) = V_0$, and therefore $\dot{W}=0$. Moreover, in the steady state the covariance matrix must be constant, and in particular $\text{d}\sigma_{PP}/\text{d}t  = 0$. On this account, the first term of \eqref{eq:current_cov} vanishes, hence the static local heat currents read as
\begin{align}
	\dot{Q}_\alpha = \tr\left[ \Pi_\alpha V_0 \sigma_{XP} M^{-1}\right]. 
\end{align}
It remains to substitute for $\sigma_{XP}$. To this aim, we note that in the spectral domain, the steady state solution is given by $X(\omega) = G_0(\omega) B(\omega)$. We also benefit from the fluctuation-dissipation relation, to arrive at \begin{align}
\dot{Q}_\alpha &= - \int_{\mathbb{R}} \frac{\text{d} \omega}{(2\pi)^2} \hbar\omega \pi \text{Im}\left\{\tr\left[\Pi_\alpha V_0 G_0(\omega) \nu(\omega) G_0^\dagger (\omega) \right]\right\},\nonumber\\
&= -\sum_\beta \int_{\mathbb{R}}\text{d} \omega \hbar\omega \text{Im}\left\{\tr\left[\Pi_\alpha V_0 G_0(\omega) J_\beta (\omega) G_0^\dagger (\omega) \right]\right\}\left(n_\beta(\omega)+\frac{1}{2}\right), \nonumber\\
&= -\sum_\beta \int_\mathbb{R} \text{d}\omega \mathcal{T}^{0}_{\alpha\beta}(\omega) \left(n_\beta(\omega)+\frac{1}{2}\right),\label{eq:static_qalpha1}
\end{align}
The equation of motion for $G_0(\omega)$ can be rearranged as $ V_0 G_0(\omega) = \mathds{1} +\omega^2M G_0(\omega) + \chi(\omega) G_0(\omega)$. For $\beta \neq \alpha$, the only non-zero contribution to the currents is given by the term proportional to the susceptibility. Using $\text{Im}\chi(\omega) = \pi J(\omega)$ we obtain $\mathcal{T}^{0}_{\alpha\beta}(\omega) = \hbar \omega \pi \tr [J_\alpha(\omega)G_0(\omega)J_\beta(\omega)G_0^\dagger(\omega)]$ for $\beta \neq \alpha$. The conservation equation $\sum_\alpha \dot{Q}_\alpha = 0$ translates into $\sum_{\beta}\mathcal{T}^0_{\alpha\beta}(\omega)=0$ and therefore we can compute $\mathcal{T}^{0}_{\alpha\alpha}(\omega)= -\sum_{\beta\neq\alpha}\mathcal{T}^{0}_{\alpha\beta}(\omega)$. Using this result into Eq.~\eqref{eq:static_qalpha1}, one gets 
\begin{align}
\dot{Q}_\alpha &=   - \int_\mathbb{R} \text{d}\omega \mathcal{T}^{0}_{\alpha\alpha}(\omega) \left(n_\alpha(\omega)+\frac{1}{2}\right)-\sum_{\beta\neq\alpha} \int_\mathbb{R} \text{d}\omega \mathcal{T}^{0}_{\alpha\beta}(\omega) \left(n_\beta(\omega)+\frac{1}{2}\right),\nonumber\\
&= \sum_{\beta\neq\alpha}\int_\mathbb{R} \text{d}\omega \mathcal{T}^{0}_{\alpha\beta}(\omega)(n_\alpha(\omega)-n_\beta(\omega))\label{eq:static_qalpha2}.
\end{align}
which is the expression shown in the main text. 

\subsection{Periodically Driven Heat Currents and Work Rate}

Let us consider the integrodifferential super-operator $\mathcal{L}_t[\blacksquare] = M (\text{d}/\text{d}t)^2 \blacksquare + V(t) \blacksquare- \chi(t)\star \blacksquare$ associated to the homogeneous quantum Langevin equation. The equation of motion of the Green's function is given by $\mathcal{L}_t [G(t,t')] = \mathds{1}\delta(t-t')$. If the potential is $\tau-$periodic, that is $V(t+\tau)=V(t)$,  and $\mathcal{L}_t[G(t,t')] = \mathds{1}\delta(t-t')$  then $ \mathcal{L}_t[G(t'+\tau,t+\tau)]=\mathds{1}\delta(t-t')$ and therefore $G(t,t') = G(t+\tau,t'+\tau)$ in the steady state by uniqueness of the solution. Now, we define the periodic function $P(t,\omega) = e^{\text{i} \omega t} \int_\mathbb{R}\text{d}t' e^{-\text{i}\omega t'} G(t,t')$. It is easy to see that $P(t+\tau, \omega)= P(t,\omega)$. Since $\mathcal{L}_t$ is a linear superoperator, we have $ \int_\mathbb{R}\text{d}t' \exp(-\text{i}\omega t')\mathcal{L}_t[G(t,t')] = \mathcal{L}_t[\exp(-\text{i}\omega t)P(t,\omega)]$ and using Eq.~(11) of the main text, we have $ \sum_k \mathcal{L}_t[\exp(-\text{i}(\omega-k\omega_d) t)]  A_k(\omega)= \mathds{1}\exp(-\text{i}\omega t)$. It follows
\begin{align}
 \sum_k  \left[ -M(\omega-k\omega_d)^2 e^{\text{i} k \omega_d t} + \sum_j V_j e^{\text{i}(j+k)\omega_d t} +\chi(\omega-k\omega_d) e^{\text{i} k\omega_d t}    \right] A_k(\omega)= \mathds{1},\label{eq:eom_Ak}
\end{align}
and projecting on $k$th element of the Fourier expansion, one arrives at 
\begin{align}
G_0^{-1}(\omega-k\omega_d) A_k(\omega) + \sum_{j\neq 0} V_j A_{k-j}(\omega) = \delta_{k,0}\mathds{1}.\label{eq:eom_Ak2}
\end{align}
We can use now the amplitudes $\{A_k(\omega)\}$ to compute the steady state solution for
\begin{align}
X(t) &= \int_\mathbb{R} \text{d}t' G(t,t') B(t') = \sum_k \int_\mathbb{R}\frac{\text{d}\omega}{2\pi} e^{-\text{i} (\omega -k \omega_d)t} A_k(\omega) B(\omega),\label{eq:position} \\
\dot{X}(t) &= \frac{d}{dt} \int_\mathbb{R} G(t,t') B(t') = -\text{i}  \sum_k \int_\mathbb{R}\frac{\text{d}\omega}{2\pi} (\omega-k\omega_d) e^{-\text{i} (\omega -k \omega_d)t}A_k(\omega) B(\omega).\label{eq:velocity}
\end{align}
The covariance matrix $\Sigma$ can be computed also using $A_k(\omega)$ and it turns out to be $\tau-$periodic. Let us take for instance 
\begin{align}
\sigma_{XP}M^{-1} =\sum_{ j k} \left[\text{i}\sum_\alpha \int_{\mathbb{R}} \text{d}\omega  \hbar(\omega- k\omega_d) A_j(\omega) J_\alpha(\omega) A^\dagger_k(\omega)\left(n_\alpha(\omega)+\frac{1}{2}\right) \right] e^{\text{i} (j-k)\omega_d t}.\label{eq:xp_cov}
\end{align}
We now take the average over a period of the heat currents in Eq.~\eqref{eq:current_cov}. Notice that since $\sigma_{PP}$ is $\tau-$periodic, the average of its derivative over a period vanishes. Then, $\bar{\dot{Q}}_\alpha = \tr \left[\Pi_\alpha \overline{V(t)\sigma_{XP} M^{-1}}\right]$ and using Eq.~(13) of the main text and Eq.~\eqref{eq:xp_cov} leads to
\begin{align}
\bar{\dot{Q}}_\alpha = -\sum_\beta \int_\mathbb{R} \text{d}\omega \mathcal{T}_{\alpha\beta}(\omega)\left(n_\beta(\omega)+\frac{1}{2}\right)=-\sum_\beta  \int_{\mathbb{R}} \text{d}\omega \sum_{k} \hbar(\omega- k\omega_d) \text{Im}\left\{ \tr \left[\Pi_\alpha \sum_{j}V_{j}  A_{k-j}(\omega) J_\beta(\omega) A^\dagger_k(\omega)\right]\right\} \left(n_\beta(\omega)+\frac{1}{2}\right).\label{av_qalpha1}
\end{align}
From Eq.~\eqref{eq:eom_Ak2} follows $\sum_j V_j A_{k-j} = \delta_{k,0}\mathds{1} + (\omega-k\omega_d)^2M A_k(\omega)+ \chi(\omega-k\omega_d) A_k(\omega) $ which can be used to compute $\mathcal{T}_{\alpha\beta}(\omega) = \sum_k \hbar(\omega-k\omega_d)\pi \tr\left[J_\alpha(\omega-k\omega_d)A_k(\omega) J_\beta (\omega) A^\dagger_k (\omega)\right]$ for $\beta\neq\alpha$. The diagonal terms can be computed using $\tilde{\mathcal{T}}_\alpha(\omega) = \mathcal{T}_{\alpha\alpha}(\omega) + \sum_\beta \mathcal{T}_{\beta\alpha}(\omega)$ where $\tilde{\mathcal{T}}_\alpha$ is given by Eq.~(16) of the main text, since $\sum_\beta \sum_{jk} \hbar\omega \text{Im}\{\tr[\Pi_\beta V_{k-j} A_j(\omega) J_\alpha(\omega) A^\dagger_k(\omega)]\} = 0$. Finally, 
\begin{align}
\bar{\dot{Q}}_\alpha = -\int_\mathbb{R}\text{d}\omega \tilde{\mathcal{T}}_\alpha(\omega) \left(n_\alpha(\omega)+\frac{1}{2}\right) + \sum_{\beta \neq \alpha} \int_\mathbb{R} \text{d}\omega\left[ \mathcal{T}_{\beta\alpha}(\omega)\left(n_\alpha(\omega)+\frac{1}{2}\right) - \mathcal{T}_{\alpha\beta}(\omega)\left(n_\beta(\omega)+\frac{1}{2}\right)\right],
\end{align}
and using the first law
\begin{align}
\bar{\dot{W}} = -\sum_\alpha \bar{\dot{Q}}_\alpha = \sum_\beta \int_\mathbb{R} \text{d}\omega \sum_\alpha \mathcal{T}_{\alpha\beta}(\omega)\left(n_\beta(\omega)+\frac{1}{2}\right) = \sum_\beta \int_\mathbb{R} \text{d}\omega\tilde{\mathcal{T}}_\beta(\omega) \left(n_\beta(\omega) +\frac{1}{2}\right),
\end{align}
which are the expressions given in the main text. 

\section{Equivalence of the Heat Currents}

The amount of heat from the bath $\alpha$ to the system S is equal to the energy lost by that reservoir, i.e. $\Delta Q_\alpha' = -(\qav{H_{R_\alpha}}_t-\qav{H_{R_\alpha}}_{t_0})$. Then, the heat current can be cast as
\begin{align}
\dot{Q}'_\alpha = -\tr\left[{\dot{H}_{R_\alpha} \rho(t)} \right] - \tr\left[{H_{R_\alpha} \dot{\rho}(t)} \right] = \tr\left[(H_S+H_{SR_\alpha})\dot{\rho}(t)\right] = \dot{Q}_\alpha - \delta\dot{Q}_\alpha,
\end{align}
where, $\tr\left[H(t)\dot{\rho}(t)\right] =0$ has been used, and $\delta\dot{Q}_\alpha$ is the difference between the two definitions of the heat currents mentioned in the main text. Equivalently, $\dot{Q}'_\alpha = -\text{i}/\hbar \left< [H_{\text{SR}_\alpha} ,H_{\text{R}_\alpha} ] \right>$ and $\dot{Q}_\alpha = \text{i}/\hbar \left< [H_{\text{SR}_\alpha} ,H_{S} ] \right>$. Then, 
\begin{align}
 \delta \dot{Q}_\alpha  =  \dot{Q}_\alpha-\dot{Q}'_\alpha = \frac{\text{i}}{\hbar} \left<[H_{\text{SR}_\alpha},H_\text{S}+H_{\text{R}_\alpha}] \right> = -\frac{\text{i}}{\hbar} \qav{[H, H_{SR_\alpha}]}.
\end{align}
Using the Heisenberg equation of motion for $H_{RS_\alpha}$ on the RHS of Eq.(4), it follows
\begin{align}
\delta \dot{Q}_\alpha =  \frac{\text{d}}{\text{d}t} \left< X^T C_\alpha X_\alpha \right> = \frac{\text{d}}{\text{d}t} \left< X^T \Pi_\alpha (\dot{P} + V(t) X)\right>. \label{eq:deltaq}
\end{align}
We are interested in $\tau$-periodic time-dependent couplings. Using Eq.~\eqref{eq:position} and Eq.~\eqref{eq:velocity} we see that, at the steady state
\begin{align}
&\left< X^T \Pi_\alpha \dot{P} \right> =\sum_{k,k'} \left\{ -  \sum_\beta \int_{\mathbb{R}} \text{d}\omega \hbar(\omega-k\omega_d)^2  \tr \left[\Pi_\alpha M A_k'(\omega) J_\beta(\omega) A_{k'}^\dagger(\omega) \right] (n_\beta(\omega)+1/2)\right\} e^{\text{i} (k-k')\omega_d t},\\
& \left< X^T \Pi_\alpha V(t) X \right> = \sum_{j,k,k'} \left\{ \sum_\beta \int_{\mathbb{R}} \text{d}\omega \hbar\, \tr\left[ \Pi_\alpha V_j A_k(\omega) J_\beta(\omega) A^\dagger_{k'}(\omega)\right](n_\beta(\omega)+1/2)\right\}e^{\text{i}(j+k-k')\omega_d t}. 
\end{align}
and, therefore, the right-hand-side of Eq.~\eqref{eq:deltaq} is the time-derivative of a $\tau$-periodic function, whose contribution over a period averages to zero.

\subsection{A Static Work Reservoir}

Heat rectification in dynamical systems its a consequence of two facts: (i) the work injected/extracted into/from the system is a useful resource to redistribute energy and (ii) periodically driven systems exhibit new energy exchange processes that have no analog in static systems. Notice, that (i) can be also achieved by introducing a third \textit{work}-reservoir that provides the energy while keeping the set up static. Let us denote $T_3$ and $n_3(\omega)$ the temperature and thermal occupation number of such work reservoir. The static currents (Eq. (9) of the manuscript) in direct and reversed configuration read 
\begin{align}
\dot{Q}_1 = \int_{\mathbb{R}} \text{d}\omega \mathcal{T}^0_{12}(\omega) (n_1(\omega)-n_2(\omega)) +\int_{\mathbb{R}} \text{d}\omega \mathcal{T}^0_{13}(\omega) (n_1(\omega)-n_3(\omega)),\\
\dot{Q}_1^r = \int_{\mathbb{R}} \text{d}\omega \mathcal{T}^0_{12}(\omega) (n_2(\omega)-n_1(\omega)) +\int_{\mathbb{R}} \text{d}\omega \mathcal{T}^0_{13}(\omega) (n_2(\omega)-n_3(\omega)).
\end{align}
We now single out the current of the third reservoir and denote it by 
\begin{align}
\dot{W} \coloneqq \dot{Q}_3 = \int_{\mathbb{R}} \text{d}\omega \mathcal{T}^0_{31}(\omega) (n_3(\omega)-n_1(\omega))+\int_{\mathbb{R}} \text{d}\omega \mathcal{T}^0_{32}(\omega) (n_3(\omega)-n_2(\omega)).
\end{align}
The first law can be now written as $\dot{Q}_1 + \dot{Q}_2+ \dot{W} = 0$. Again, we can think about the local contributions to reservoir $\alpha = 1,2$ as $\dot{W}_\alpha = \int_{\mathbb{R}} \text{d}\omega \mathcal{T}^0_{ 3 \alpha}(\omega) (n_3(\omega)-n_\alpha(\omega))$. It is clear, that $R(\dot{Q}_1,\dot{Q}_1^r) \propto | \dot{W}_1 + \dot{W}_1^r| \neq 0$ under the interchange $T_1 \leftrightarrow T_2$. However, there is no genuine asymmetric transport happening in the sense that $\mathcal{T}^0_{12}(\omega)=\mathcal{T}^0_{21}(\omega)$ for a static system (as explained in the manuscript). By defining $\dot{\mathbb{Q}}_\alpha\coloneqq \dot{Q}_\alpha + \dot{W}_\alpha$, we only take into account the genuine contributions to the rectification, such that, in the static case $R(\dot{\mathbb{Q}}_\alpha,\dot{\mathbb{Q}}_\alpha^r)=0$.  With this intuition, we use $R(\dot{\mathbb{Q}}_\alpha,\dot{\mathbb{Q}}_\alpha^r)$ as a measure of the asymmetric transport in the driven network that is present \emph{exclusively} due to periodic forcing. 

\subsection{Static Green's Function for $N=2$}

We now give a method to obtain an analytic expression for the static Green's function $G_0(\omega)$ for the scenario of $N=2$. Namely, we assume a homogeneous network of two oscillators with $m_1 = m_2 = 1$ and coupled to identical reservoirs, in such a way that $\chi(\omega) = \tilde{\chi} (\omega)\mathds{1}$ where $\tilde{\chi}(\omega)$ is a scalar function. Then, the Green's function can be written as 
\begin{align}
G_0(\omega) = (-\omega^2\mathds{1}+V_0-\tilde{\chi}(\omega)\mathds{1})^{-1}.
\end{align}
Since $V_0$ is a symmetric real matrix, there exists a set $\{u_i\}_{i=1}^N$ of eigenmodes such that $V_0 u_i = \nu_i^2 u_i$, where $\nu_i$ are the normal mode frequencies and therefore, $V_0 = \sum_i \nu_i^2 u_i u_i^T$. Moreover, $\mathds{1}$ can be also resolved in any basis, in particular $\mathds{1} = \sum_i u_i u_i^T$. Hence
\begin{align}
G_0(\omega) =\sum_i \frac{u_i u_i^T}{\nu_i^2-\omega^2 - \text{Re}\{\tilde{\chi}(\omega)\}-\text{i} \pi \tilde{J}(\omega)} .
\end{align}
We further assume that for large enough cutoff $\Lambda_0$, $\text{Re}\{\tilde{\chi}(\omega)\}$ is only shifting the position of the normal modes by a constant amount, which can be compensated by renormalizing the Hamiltonian with the counterterm \cite{Breuer2007}. Hence, from now on, we disregard $\text{Re}\{\tilde{\chi}(\omega)\}$. The normal mode vectors $u_i$ are real, orthogonal and normalized to one, and therefore can be parametrized with an angle $\theta$ such that $u_1 = (\sin(\theta),\cos(\theta))$ and $u_2 = (\cos(\theta),-\sin(\theta))$. The angle $\theta$ is determined as a function of the system parameters as $\tan(2\theta) = -2c_0/( \omega_2^2-\omega_1^2)$. Then, for arbitrary normal modes
\begin{align}
G_0(\omega) = \left(\begin{matrix}
\frac{\cos^2(\theta)}{\nu_1^2-\omega^2-\text{i} \pi \tilde{J}(\omega)}+
\frac{\sin^2(\theta)}{\nu_2^2-\omega^2-\text{i} \pi \tilde{J}(\omega)}   && \sin(\theta)\cos(\theta) \left( \frac{1}{\nu_1^2-\omega^2-\text{i} \pi \tilde{J}(\omega)} -\frac{1}{\nu_1^2-\omega^2-\text{i} \pi \tilde{J}(\omega)}\right)\\
\sin(\theta)\cos(\theta) \left( \frac{1}{\nu_1^2-\omega^2-\text{i} \pi \tilde{J}(\omega)} -\frac{1}{\nu_1^2-\omega^2-\text{i} \pi \tilde{J}(\omega)}\right) && \frac{\sin^2(\theta)}{\nu_1^2-\omega^2-\text{i} \pi \tilde{J}(\omega)}+
\frac{\cos^2(\theta)}{\nu_2^2-\omega^2-\text{i} \pi \tilde{J}(\omega)}
\end{matrix}\right). \label{eq:greens2}
\end{align}

\subsection{Main Contributions to Rectification}

The driven harmonic network under consideration is very general and, therefore, it is rather complicated to picture the rectification coefficient as a function of the network parameters. Here, we propose to take advantage of the peaked structure of $G_0(\omega)$ obtained above, and consider only the the principal contributions to the asymmetric transport in frequency domain. We start using Eq.~(19) of the main text to write 
\begin{align}
\mathcal{T}_{21}(\omega) -\mathcal{T}_{12}(\omega) = \sum_k \hbar (\omega-k\omega_d) J(\omega-k\omega_d) J(\omega) \pi \text{v}_1^2\left( |G_0(\omega-k\omega_d)_{21}G_0(\omega)_{11}|^2-|G_0(\omega-k\omega_d)_{11}G_0(\omega)_{12}|^2\right).
\end{align}
Therefore, highly asymmetric heat transport can only occur when $|G_0(\omega-k\omega_d)_{21}G_0(\omega)_{11}|^2$ and $|G_0(\omega-k\omega_d)_{11}G_0(\omega)_{12}|^2$ are large compared with the static transport. We have seen that the Green's function $G_0(\omega)$ is peaked at $\omega = \pm \nu_i$. Hence, the principal contributions to the asymmetric transport will occur when we hit the resonances of both Green's functions at once, that is, when $k \omega_d = \pm \nu_i \pm \nu_j$. However, notice that when $i=j$ we are evaluating $|G_0(\nu_i)_{21}G_0(\nu_i)_{11}|^2-|G_0(\nu_i)_{11}G_0(\nu_i)_{12}|^2 = 0$ and consequently rectification will be small.  Therefore, only at $k \omega_d = \pm \nu_i \pm \nu_j$ with $i\neq j$ the asymmetric heat transport is expected to be large. In the particular, if we go back to the case of $N=2$, the positive driving frequencies at which one expects large rectification are $\omega_d = \nu_2 \pm \nu_1$.

\subsection{Harmonic Bipolar Transistors}
Let us consider a static bipolar thermal transistor where the third energy source is provided by a third reservoir, that is $\dot{E}=\dot{Q}_3$. The control parameter is chosen to be $T_3$, the temperature of the third bath. If we keep the rest of the variables constant, we have 
\begin{align}
\frac{\text{d}\dot{Q}_1}{\text{d}\dot{Q}_3} = \frac{\partial \dot{Q}_1/\partial T_3}{\partial \dot{Q}_3/\partial T_3}.
\end{align}
We already know the dependence of such currents in $T_3$, given in Eq.~(9) of the main text. Then, we can easily compute
\begin{align}
&\frac{\partial \dot{Q}_{\alpha}}{\partial T_3} = - \int_\mathbb{R} \text{d}\omega \mathcal{T}^{(0)}_{\alpha3} \frac{\partial n_3(\omega)}{\partial T_3} = -I_{\alpha 3}, \\
&\frac{\partial \dot{Q}_3}{\partial T_3} = \int_\mathbb{R} \text{d}\omega \left(\mathcal{T}^{(0)}_{13}(\omega)+\mathcal{T}^{(0)}_{23}(\omega)\right) \frac{\partial n_3(\omega)}{\partial T_3} = I_{13} + I_{23}.
\end{align}
Since $n_3(\omega)$ increases monotonically with $T_3$, we have $I_{12}$, $I_{23} \geq 0$. Therefore, for $\alpha=1,2$ we have $|\mathcal{A}_{\alpha}|\leq 1$ regardless of any other system parameter. 
 
\end{document}